\documentclass[preprint,amsmath,amssymb,aps,a4]{revtex4-2}
\usepackage{graphicx}
\usepackage{lineno}
\usepackage{bm}
\usepackage{txfonts}
\usepackage{hyperref}
\usepackage{caption}
\usepackage{amsmath}
\usepackage{amssymb}
\usepackage[section]{placeins}
\usepackage{tabularx}
\usepackage{braket}
\usepackage{multirow}
\usepackage{subcaption}
\usepackage{rotating}
\usepackage{xcolor}
\usepackage{cancel}
\usepackage{ulem}
\date{\today}
\newcommand{\be}{\begin{eqnarray}}
\newcommand{\ee}{\end{eqnarray}}

\begin{document}
\title{Impact of isospin asymmetric nuclear medium on pseudoscalar and vector $B$ mesons}
\author{Tanisha$^{1}$}
\email{tanisha220902@gmail.com}
\author{Satyajit Puhan$^{1}$}
\email{puhansatyajit@gmail.com}
\author{Navpreet Kaur$^{1}$}
\email{knavpreet.hep@gmail.com}
 \author{Arvind Kumar$^{1}$}
\email{kumara@nitj.ac.in}
\author{Harleen Dahiya$^{1}$}
\email{dahiyah@nitj.ac.in}

\affiliation{$^1$ Computational High Energy Physics Lab, Department of Physics, Dr. B.R. Ambedkar National
	Institute of Technology, Jalandhar, 144008, India}

\date{\today}%
\begin{abstract}
In this study, using the light-front quark model, we examine how an isospin asymmetric nuclear medium affects the properties of pseudoscalar ($B^+, B^0$) and vector ($B^{\ast+}, B^{\ast0}$) mesons under different temperature values and degrees of isospin asymmetry. To simulate the in-medium modifications of the constituent quark masses, we employ the chiral SU(3) quark mean field model. 
Our analysis focuses on evaluating the effective masses, weak decay constants, and distribution amplitudes (DAs) of $B$ mesons in isospin-asymmetric nuclear matter. The calculated vacuum values of the $B$ meson masses and decay constants show good agreement with existing experimental data, validating our approach to study the medium effects within the same framework.

\vspace{0.1cm}
\end{abstract}
%
\maketitle
%
%

\section{Introduction\label{secintro}}
One of the key challenges of the present times in hadronic and nuclear physics is uncovering how the hadrons behave when immersed in a nuclear medium. The interaction of a hadron with the medium significantly alters the internal structure of hadrons and affects meson photoproduction in heavy-ion collisions \cite{NA49:2007stj}. These modifications also play a crucial role in exploring properties of quantum chromodynamics (QCD), including chiral symmetry breaking at low energies and its restoration at higher densities and temperatures. Quarks and gluons are the fundamental degrees of freedom in QCD, which is the theory of strong interactions.  The internal dynamics of quarks and gluons, which are bound within hadrons due to color confinement, reveal important information about the vacuum properties of hadrons. The strong coupling constant \( \alpha_s \) in QCD is much larger than the fine-structure constant \( \alpha \) in quantum electrodynamics, particularly at low energies, leading to the phenomenon of confinement, where quarks and gluons cannot exist freely. Due to this strong coupling at the low scale, quark dynamics is studied using effective field theories, while high-energy processes are accessible via perturbative methods because of asymptotic freedom.\par
In a low-energy regime (\(\textit{Q}^{2} \leq\) 1 GeV\(^2\)), the non-perturbative aspects can be investigated through the valence quark distribution functions. These functions can subsequently be used as inputs to evolve the distributions, leading to perturbative QCD phenomena that are experimentally observable at higher \( \textit{Q}^{2}\). Experimental studies have shown that the nucleon structure function $F_2 (x)$ undergoes modification within nuclei, a phenomenon referred to as the European Muon Collaboration (EMC) effect \cite{EuropeanMuon:1983wih},  observed in deep inelastic scattering experiment. It showed that in the intermediate longitudinal momentum fraction $x$ range (0.3 – 0.7), the per-nucleon structure function is suppressed in heavier nuclei compared to deuterium, indicating altered quark distributions in the nuclear environment. Prior to the EMC experiment, SLAC's (Stanford Linear Accelerator Center)  deep inelastic scattering studies on nuclei like deuterium, carbon, iron, and gold also revealed scaling violations and indicated nuclear dependence in structure functions \cite{Friedman:1972sy}. Jefferson Lab (JLab) continues to study the EMC effect with high precision, focusing on light nuclei \cite{Seely:2009gt}. The MARATHON experiment \cite{Abrams:2024wgt} explores structure functions in mirror nuclei, while other studies examine nucleon correlations and their impact on quark distributions. These efforts deepen our understanding of nuclear structure and quark-gluon interactions. 

Consequently, hadron characteristics are projected to undergo modifications in the nuclear medium related with partial restoration of chiral symmetry. Experimental evidence for the partial restoration of chiral symmetry has been supported by observations in deeply bound pionic atoms \cite{Suzuki:2002ae}, low-energy pion-nucleus scattering \cite{Friedman:2004jh}, and di-pion production in hadron-nucleus and photon-nucleus reactions \cite{CHAOS:1996nql, CHAOS:2004rhl}. Theoretical investigations of hadron modification in nuclear medium, particularly for heavy and light mesons, have been extensively conducted across a variety of models and approaches, such as the quark-meson coupling (QMC) model \cite{Saito:2005rv, Krein:2017usp, Zeminiani:2020aho, Cobos-Martinez:2022fmt},  the Bethe-Salpeter equation-Nambu-Jona-Lasinio (BSE-NJL) model \cite{Hutauruk:2018qku, Hutauruk:2021kej, Hutauruk:2019ipp, Hutauruk:2019was}, the QCD sum rule (QSR) \cite{Park:2016xrw, Bozkir:2022lyk}, the holographic model \cite{Kim:2022lng},  the Dyson-Schwinger equation (DSE) based approach \cite{Roberts:2000aa}, the Linear-sigma model (L$\sigma$M) \cite{Suenaga:2019urn}, the instanton liquid model (ILM) \cite{Nam:2008xx, Shuryak:1997vd}, and the hybrid light-front quark-meson coupling (LF-QMC) models \cite{deMelo:2016uwj,deMelo:2014gea,deMelo:2018hfw}.
\par
Heavy-ion collision experiments in the Relativistic Heavy Ion Collider (RHIC) at Brookhaven National Laboratory and the Large Hadron Collider (LHC) use heavy quarks like charm and bottom as valuable probes to explore nuclear matter under extreme temperatures and lower baryonic densities. In contrast, experiments at higher baryonic densities, such as BM@N (Baryonic Matter at Nuclotron) at the NICA accelerator complex at the Joint Institute for Nuclear Research (JINR), are aimed to focus on the reaction dynamics and phase transitions of dense matter \cite{Afanasiev:2025kbf}. Facilities like the Schwerionen Synchrotron (SIS) accelerator at GSI \cite{FOPI:2010xrt} and the upcoming Compressed Baryonic Matter (CBM) experiment at FAIR \cite{CBM:2016kpk, Durante:2019hzd, Agarwal:2022ydl} explore properties of strongly interacting hot and dense nuclear matter.
Given recent advancements, it becomes desirable to investigate the properties of $B$ mesons, both in vacuum and nuclear medium. These mesons, consisting of a lighter quark (antiquark), which can be \(c, s, d\) or \( u\), bound to a heavy $b$ antiquark (quark)  present a unique opportunity to explore their behavior in different environments. The decays of $B$ mesons are utilized for precise measurements of the Cabibbo-Kobayashi-Maskawa (CKM) matrix elements \cite{Duplancic:2008zz} and for testing CP violation within the Standard Model \cite{Kobayashi:1973fv}. Considerable advancements have been made in the study of $B$ mesons over the past few years. $B$ mesons are unique, as they are the only mesons that contain third-generation quarks. The heavy \(b\) quark in $B$ mesons opens up numerous possible decay channels \cite{Browder:1995gi}. However, the medium modification of \(B\) mesons arises from the interaction between the lighter quark and the medium, while the degree of freedom associated with the bottom quark is considered effectively frozen in the medium due to its significantly larger mass. Even though there is considerable interest in $B$ mesons and their properties in nuclear matter, the available literature on $B$ mesons in medium is relatively limited. Different methods, such as the QCD sum rule and the QMC model, have investigated how the properties of \(B\) mesons—such as masses and decay constants—are altered in both symmetric and asymmetric nuclear \cite{Tsushima:2002cc, Dhale:2018plh, Yasui:2012rw, Hilger:2008jg, Wang:2015uya, Azizi:2014bba}, and strange hadronic medium \cite{Pathak:2014nfa, Chhabra:2016vhp}. Furthermore, the transport models have explained how \(B\) mesons interact with the medium by incorporating drag and diffusion coefficients, resulting in changes to their momentum and energy distributions \cite{Abreu:2012et}. \par
This study seeks to enhance the understanding of hadron-nucleus interactions in a nuclear medium by examining the changes in the effective masses of pseudoscalar and vector $B$ mesons ($B^{+}, B^{0}, B^{\ast+}, B^{\ast0}$) in an isospin asymmetric nuclear enviroment. We have also calculated the medium-modified weak decay constants and distribution amplitudes (DAs) at finite temperature T and isospin asymmetry \(\eta\). Mass modifications in the nuclear medium, attributed to variations in the scalar mean field in certain models, are often associated with the quark chiral condensate, indicating the partial restoration of chiral symmetry. The weak decay constants describe the strength of decays of hadrons through weak interactions \cite{Buchalla:1995vs}.  In particular, the weak decay constants of the heavy mesons play a crucial role in studying CP violation \cite{Blanke:2017ohr}, CKM matrix elements, $B-\bar{B}$ mixing, and leptonic or nonleptonic weak decay.
The DAs of mesons provide important insights into the non-perturbative structure of hadrons and the distribution of partons, particularly in terms of their longitudinal momentum fractions within these particles \cite{Lepage:1980fj}.\par 

In this work, we have employed a combined approach of light-front quark model (LFQM) and chiral SU(3) quark mean field (CQMF) model to analyze the in-medium properties of pseudoscalar ($B^{+}, B^{0}$) and vector ($B^{\ast+}, B^{\ast0}$) mesons at finite baryonic density ratio $\rho_B/\rho_0$, temperature T, and isospin asymmetry $\eta$. The vacuum properties of $B$ mesons are derived using the LFQM, and the in-medium effects are incorporated through CQMF. The LFQM is a theoretical framework based on light-front dynamics, which has proven highly effective in explaining both the mass spectra and wave function related observables, including the electroweak properties of mesons, such as form factors (FFs), charge radii, parton distribution functions (PDFs), decay widths, etc. The light-front wave functions (LFWFs) are the Fock-state projections of the eigenstates of the QCD light-front Hamiltonian. These LFWFs are boost-invariant, which means they are independent of the hadron’s longitudinal momentum \( P^+ = P^0 + P^z \) and transverse momentum \( \textbf{\textit{P}}_{\perp}\). The LFWFs can be expressed using internal momentum fraction variables that are independent of the hadron's momentum, thus ensuring explicit Lorentz invariance \cite{Brodsky:2001wx}. The examination of mass spectra and decay constants is conducted by establishing model parameters and employing a trial wave function for the variational principle applied to a QCD-inspired Hamiltonian that includes the standard kinetic term, Coulomb potential, along with linear confining potential, and a perturbative hyperfine interaction potential. In the CQMF model, quarks are treated as degrees of freedom, confined within baryons by a confining potential. Through the interchange of vector fields \( \omega \) and \( \rho \), as well as scalar fields \( \sigma \), \( \zeta \), and \( \delta \), these quarks interact inside the baryons. For the scalar meson with strange quark content, the \(\zeta\) field is considered, which is termed as strange scalar-isoscalar field. The mean-field relativistic models incorporate the vector-isovector field \( \rho \) and the scalar isovector field \( \delta \) to account for the isospin asymmetry of the medium. The dilaton field \(\chi\) is also incorporated in this model to capture the broken scale invariance of QCD \cite{Wang:2004wja}. This model has also been used to examine the magnetic moments of octet and decuplet baryons in nuclear medium \cite{Singh:2016hiw, Singh:2020nwp} and strange hadronic medium \cite{Singh:2018kwq, Singh:2018tdm}.
From the CQMF model, we determine the medium modified effective masses of the constituent quarks of the mesons, which will then be used as input in LFQM to study the modifications in the properties of $B$ mesons. 
\par The outline of the paper is as follows. In Sec. \ref{sec2a}, we provide the description of the LFQM approach by introducing the effective Hamiltonian and LFWFs. Additionally, we explain how to determine model parameters using the variational principle. In Sec. \ref{sec2b} and \ref{sec2c}, we summarize the expressions of weak decay constants and distribution amplitudes, respectively. In Sec. \ref{sec2d}, we outline the CQMF model to explore in-medium effects in LFQM calculations. In Sec. \ref{sec3}, we discuss the results of our work and summarize them in Sec. \ref{sec4}.
\section{Methodology }
\subsection{Light-front quark model } \label{sec2a}
The LFQM provides a suitable framework to describe the hadrons in terms of their constituent quarks.
The meson bound state having a quark (q) antiquark ($\bar q$) pair can be expressed in terms of four vector momenta $\textbf{P}=(\textbf{P}^{+}, \textbf{P}^{-},\textbf{P}_{\perp})$, total angular momentum \((J, J_z)\), and in the form of a meson eigenstate $|\mathbf{\mathcal{M}}(\textbf{\textit{P}}, J, J_z)\rangle$ as \cite{Arifi:2023jfe, Choi:1997iq}
\begin{equation}
|\mathbf{\mathcal{M}}(\textbf{\textit{P}}, J, J_z)\rangle = \int \left[\mathrm{d}^3\mathbf{p}_q\right] \left[\mathrm{d}^3\mathbf{p}_{\bar{q}}\right] 2(2\pi)^3 \delta^3(\mathbf{P} - \mathbf{p}_q - \mathbf{p}_{\bar{q}})
\times \sum_{\lambda_q, \lambda_{\bar{q}}} \mathbf{\Psi}^{J J_z}_{\lambda_q \lambda_{\bar{q}}}(x, \mathbf{k}_\perp) |q_{\lambda_q}(p_q) \bar{q}_{\lambda_{\bar{q}}}(p_{\bar{q}})\rangle,\end{equation}
where \(\mathbf{p}_j = (p^{+}_{j}, \mathbf{p}_{\perp} )\) and \(\left[\mathrm{d}^3\mathbf{p}_i\right] \equiv \mathrm{d}p^{+}_{j} \mathrm{d}^2\mathbf{p}_{j\perp}/[2(2\pi)^3]
 \). The momentum and helicity of a quark (antiquark) \(j=q, \bar{q}\) are represented as \(p_{q(\bar{q})}\) and \(\lambda_{q(\bar{q})}\), respectively. The internal quark variables of the light-front \( (x, \mathbf{k}_\perp) \) are defined as \(x = {p^+_q}/\mathbf{P}^+\) being the longitudinal momentum fraction and \(\mathbf{k}_\perp = \mathbf{p}_{q \perp} - x \mathbf{P}_\perp \) is the transverse momenta of the active quark.
 In momentum space, the LFWF $\mathbf{\Psi}^{J J_z}_{\lambda_q \lambda_{\bar{q}}}$ of the ground state meson with different forms of quark antiquark helicities is defined by
 \begin{equation}
    \mathbf{\Psi}^{J J_z}_{\lambda_q \lambda_{\bar{q}}}(x, \mathbf{k}_\perp) = \mathbf{\Phi}(x, \mathbf{k}_\perp) \mathbf{\mathcal{R}}^{J J_z}_{\lambda_q \lambda_{\bar{q}}}(x, \mathbf{k}_\perp),
    \label{wavefn}
    \end{equation}
where \(\mathbf{\Phi}(x, \mathbf{k}_\perp)\) is the radial wave function and the spin-orbit wave function is denoted by \(\mathbf{\mathcal{R}}^{JJ_z}_{\lambda_q\lambda_{\bar{q}}} (x, \mathbf{k}_\perp)\) which distinguishes the vector (V) and pseudoscalar (P) mesons, is obtained by the interaction-independent Melosh transformation \cite{Melosh:1974cu}.
The covariant forms of \(\mathbf{\mathcal{R}}^{JJ_z}_{\lambda_q\lambda_{\bar{q}}}\) is given by
\begin{equation}
    \mathcal{R}_{\lambda_{q}\lambda_{\overline{q}}}^{JJ_{z}}=\frac{1}{\sqrt{2}\tilde{\textbf{\textit{M}}}_{0}}\bar{u}_{\lambda_{q}}(p_{q})\Gamma_{M}v_{\lambda_{q}}(p_{\bar{q}}),
\end{equation}
with $M$ = $P$ or $V$ and \(\tilde{{\textbf{\textit{M}}}}_{0}=\sqrt {\textbf{\textit{M}}_0^2 - (m_q - m_{\bar{q}})^2}\), $m_{q(\bar q)}$ is the quark (antiquark) mass. The invariant meson mass squared, \(\textbf{\textit{M}}_0^2\) is given by
\begin{equation}
   \textbf{\textit{M}}_0^2= \frac{\textbf{k}^{2}_{\perp}+m^{2}_{q}}{x} + \frac{\textbf{k}^{2}_{\perp}+m^{2}_{\bar{q}}}{1-x}.
\end{equation}
Also, the vertex factor \( \mathbf{\Gamma}_M \) for pseudoscalar and vector mesons, respectively, is defined by\\
 \begin{equation}
    \mathbf{\Gamma}_P=\gamma_5,\end{equation}    
\begin{equation}    
\mathbf{\Gamma}_V = - \cancel{\epsilon}(J_z) + \frac{\epsilon \cdot (p_q - p_{\bar{q}})}{\textit{\textbf{M}}_0 + m_q + m_{\bar{q}}},\end{equation} 
where the polarization vectors  of the vector meson \( \epsilon^\mu(J_z) = (\epsilon^+, \epsilon^-, \mathbf{\epsilon}_{\perp})\) are given by
\[
\mathbf{\epsilon}^\mu(\pm 1) = 
\left( 
0, \frac{2 \, \mathbf{\epsilon}_\perp(\pm) \cdot \mathbf{P}_\perp}{P^+}, \mathbf{\epsilon}_\perp(\pm)
\right),
\]
\begin{equation}
    \mathbf{\epsilon}^\mu(0) = 
\frac{1}{\textbf{\textit{M}}_0} 
\left( 
P^+, \frac{-\textit{\textbf{M}}_0^2 + \mathbf{P}_\perp^2}{P^+}, \mathbf{P}_\perp
\right),
\end{equation}
with \( \mathbf{\epsilon}_\perp (\pm 1) = \pm\frac{1}{\sqrt{2}} (1,\pm i)\). The spin-orbit wave function \(\mathbf{\mathcal{R}}^{JJ_z}_{\lambda_q\lambda_{\bar{q}}}\) for both pseudoscalar and vector mesons is normalized to unity, satisfying the relations \begin{equation} \sum_{\lambda_q, \lambda_{\bar{q}}}\mathcal{R}_{\lambda_{q}\lambda_{\bar{q}}}^{JJ_{z}\dagger}\mathcal{R}_{\lambda_{q}\lambda_{\bar{q}}}^{JJ_{z}}=1.  
\end{equation}
We use the Gaussian-type wave function for the ground state of the meson as the trial radial wave function
\begin{equation}
    \mathbf{\Phi}_{1S}(x, \mathbf{k}_\perp) = \frac{4\pi^{3/4}}{\beta^{3/2}} \sqrt{\frac{\partial k_{z}}{\partial x}} e^{-\textbf{k}^{2}/2\beta^{2}},
\end{equation}
where \(\beta\) is the variational parameter that corresponds to the size of the wave function and \(\textbf{k}^{2}=\textbf{k}^2_\perp + k^2_z\) is the internal momentum of the meson. The Jacobian factor \(\partial k_{z}/\partial x\), responsible for variable transformation \((k_z, \textbf{k}_\perp) \rightarrow \)  \((x, \textbf{k}_\perp)\) is given by
\begin{equation}
    \frac{\partial k_{z}}{\partial x}=\frac{\textbf{\textit{M}}_{0}}{4 x (1-x)}\bigg[1-\frac{(m_{q}^{2}-m_{\bar q}^{2})^{2}}{\textbf{\textit{M}}^{4}_0}\bigg],
\end{equation}
where \(k_{z}=(x-1/2) \textbf{\textit{M}}_{0}+(m_{\bar q}^{2}-m_{q}^{2})/2\textbf{\textit{M}}_{0}\). As the radial wave function is also normalized in LFQM, so the LFWF also satisfies the following normalization condition
\begin{equation}
    \int \frac{\mathrm{d}x\mathrm{d}^2\textbf{k}_\perp}{2(2\pi)^3} |\mathbf{\Psi}(x,\textbf{k}_\perp)|^2 = 1.
\end{equation}
The interactions between quarks and antiquarks are incorporated into the meson mass operator to calculate the mass eigenvalue \( \textbf{\textit{M}}_{q\bar{q}} \)
\begin{equation}
     H_{q \bar{q}}|\mathbf{\Psi}_{q \bar{q}} \rangle = \textbf{\textit{M}}_{q\bar{q}} | \mathbf{\Psi}_{q \bar{q}} \rangle,
\end{equation}
where the eigenfunction for the meson is \(\mathbf{\Psi}_{q \bar{q}}\). The QCD- motivated effective Hamiltonian, \(H_{q \bar{q}}\) in the quark-antiquark center of mass frame is expressed as \(H_{q \bar{q}} = H_0 + V_{q\bar{q}}\), where \(H_0 = \sqrt{m^2_q + \textbf{k}^2} + \sqrt{m^2_{\bar{q}} + \textbf{k}^2}\) is the relativistic kinetic energy of the quark and antiquark. The effective interaction potential between a quark and an antiquark in the rest frame of the meson \(V_{q\bar {q}}\) is taken in the form of \cite{Arifi:2023jfe}
\begin{equation}
    V_{q \bar{q}}=V_{Conf} + V_{Coul} + V_{Hyp}, 
\end{equation}
where $ V_{Conf}$, $V_{Coul}$ and $V_{Hyp}$ correspond to the Coulomb potential, linear confining potential and hyperfine interaction potential, respectively. Their expressions have been written as follows \cite{Arifi:2023jfe}
\begin{equation}
    V_{Conf}=a + br,
\end{equation}
\begin{equation}
   V_{Coul}=-\frac{4\alpha_{s}}{3r},
\end{equation}
\begin{equation}
    V_{Hyp}=\frac{32\pi\alpha_{s} \langle S_q . S_{\bar q}\rangle}{9m_{q}m_{\bar{q}}} \delta^3(r), \label{hyperfine}
\end{equation}
respectively. The linear confining potential parameters are $a$ and $b$, while the strong running coupling, denoted as \(\alpha_s\), is a constant parameter in the vacuum. For vector and pseudoscalar mesons, the values of \(\langle S_q. S_{\bar q}\rangle\) are 1/4 and -3/4, respectively.\par
 Using the variational principle, we evaluate the expectation value of the Hamiltonian to compute the mass eigenvalue of meson \(\textbf{\textit{M}}_{q \bar{q}} =\langle \mathbf{\Phi}_{1S}|H_{q\bar{q}}|\mathbf{\Phi}_{1S} \rangle \) using a Gaussian trial wave function \(\mathbf{\Phi}(x, \textbf{k}_\perp)\) with the variational parameter \( \beta \). The analytic form of the mass eigenvalue is obtained as \cite{Choi:2009ai}
 \begin{equation}
     \textbf{\textit{M}}_{\bar{q}q} = \frac{\beta}{\sqrt{\pi}} \sum_{i=q,\bar{q}} z_i e^{-z_i^2/2} K_1\left(\frac{z_i}{2}\right) + a + \frac{2b}{\beta\sqrt{\pi}} - \frac{8\alpha_s \beta}{3\sqrt{\pi}} + \frac{32\alpha_s \beta^3 \langle \mathbf{S}_q \cdot \mathbf{S}_{\bar{q}} \rangle}{9\sqrt{\pi} m_q m_{\bar{q}}}, \label{17}
 \end{equation}
 where \(z_{i}=m_{i}^2/ \beta^2\) and \(K_n\) is the modified Bessel function of the second kind of order-$n$.  The variational parameter \( \beta \) is obtained by applying the variational principle and ensuring that \( \alpha_s \) is constant for all mesons. As the main objective of our work is to investigate the modifications of meson structure and characteristics in the medium, we employ the same parameters as in vacuum. We introduce a Gaussian smearing function, i.e., \(\delta^3(\textbf{r}) \rightarrow (\frac{\sigma'^{3}}{\pi^{3/2}}) e^{\sigma'^2 \textbf{r}^2}\), to weaken the singularity of the $\delta^3(\textbf{r})$ function in hyperfine interactions and to determine the true minimum value for the mass at a given value of $\beta$. This will prevent the $\sigma'$ function from generating negative infinity.
 A more detailed procedure is provided in Ref. \cite{Choi:1997iq}.
\subsection{Weak decay constants} \label{sec2b}
One of the basic features of mesons is the weak decay constant. The decay constant is defined through the matrix elements
of axial-vector and vector currents between the meson state and vacuum with 
four-vector momenta $P$ for the pseudoscalar \(f_P\) and vector \(f_V\) mesons as \cite{Arifi:2023jfe,Choi:2007se} 
\[ \langle 0| \bar{q} \gamma_{\mu} \gamma_{5} q|P(\textbf{\textit{P}}) \rangle = i f_{P} \textbf{\textit{P}}^{\mu},\]
\begin{equation}
    \langle 0| \bar{q} \gamma_{\mu} q|V(\textbf{\textit{P}}) \rangle = f_V \textbf{\textit{M}}_{V} \mathbf{\epsilon}^{\mu}(J_{z}),
\end{equation}
with \(\textbf{\textit{M}}_{V}\) being the mass of the vector meson. In LFQM, the explicit form of the decay constant for a meson $M$ (that can be a pseudoscalar $P$ or vector $V$ meson) is given by \cite{Jaus:1991cy}
\begin{equation}
    f_{M} = 2\sqrt{6} \int^{1}_{0} \mathrm{d}x \int \frac{\mathrm{d}^{2}\textbf{k}_\perp}{2(2 \pi)^3} \frac{\mathbf{\Phi}(x, \textbf{k}_\perp)}{\sqrt{\mathcal{A}^{2}+ \textbf{k}^{2}_\perp}} \mathcal{O}_{M}, \label{19}
\end{equation}
with the operator \(\mathcal{O}_{M}\) as
\[\mathcal{O}_{P}=\mathcal{A},\]
\begin{equation}   \mathcal{O}_{V}=\mathcal{A}+\frac{2 \textbf{k}_\perp^2}{\textbf{\textit{M}}_0 + m_{q} + m_{\bar{q}}}, \label{20}
\end{equation}
where \(\mathcal{A}=(1-x)m_{q}+x m_{\bar{q}}\).
\subsection{Distribution amplitude} \label{sec2c}
The leading twist quark DAs of the corresponding mesons influence the probabilities of valence quark distributions and can be obtained through the analysis of hard exclusive reaction processes \cite{Chernyak:1983ej, Lepage:1980fj}.
The DAs are defined by gauge-invariant, light-like separated matrix elements from free space to mesons and can be interpreted as probabilities to find a meson in states with a minimum number of Fock constituents and small transverse momentum separation.\par
The leading twist DAs for the pseudoscalar and vector mesons  can be derived from the positive component of the currents, which are respectively expressed as \cite{Choi:2007se,Hwang:2010hw}
\begin{align}
\textbf{\textit{A}}_P^+ &= \langle 0 | \bar{q}(z) \gamma^+ \gamma_5 q(-z) | P(\textbf{\textit{P}}) \rangle = i f_P P^+ \int_0^1 \mathrm{d}x \, e^{i \zeta \textbf{\textit{P}} \cdot z} \phi_\textbf{\textit{P}}(x) \bigg|_{z^+ = z_\perp = 0},  \\[10pt]
\textbf{\textit{A}}_{V}^{+} &= \langle 0 | \bar{q}(z) \gamma^+ q(-z) | V(\textbf{\textit{P}}, 0) \rangle= f_V \textbf{\textit{M}}_V \mathbf{\epsilon}^{+} (0) \int_0^1 \mathrm{d}x \, e^{i \zeta \textbf{\textit{P}} \cdot z} \phi_V(x) \bigg|_{z^+ = z_\perp = 0},
\end{align}
where \(\zeta = 2x-1\). The explicit form of DAs for vector and pseudoscalar mesons in LFQM can be given by 
\begin{align}
    \phi_{\mathrm{M}}(x) = \frac{2\sqrt{6}}{f_{\mathrm{M}}} \int \frac{\mathrm{d}^2 \mathbf{k}_{\perp}}{2(2\pi)^3} \frac{\mathbf{\Phi}(x, \mathbf{k}_{\perp})}{\sqrt{\mathcal{A}^2 + \mathbf{k}_{\perp}^2}} \mathcal{O}_{M}. \label{23}
\end{align}
The DAs are normalized as
\begin{align}
    \int^{1}_{0} \phi_{M}(x) \mathrm{d}x=1.
\end{align}
\subsection{ Chiral SU(3) quark mean field model} \label{sec2d}
The effective Lagrangian density of the CQMF model describing various interactions can be defined as \cite{Wang:2001jw}
\begin{align}
    \mathcal{L}_{\mathrm{eff}} = \mathcal{L}_{q0} + \mathcal{L}_{qm} + \mathcal{L}_{\Sigma \Sigma} + \mathcal{L}_{VV} + \mathcal{L}_{\chi SB} + \mathcal{L}_{\Delta m} + \mathcal{L}_{c}. \label{L}
\end{align}
The first term represents the kinetic term for free massless quarks, given by \( \mathcal{L}_{q0} = \bar{q} \, i \gamma^{\mu} \partial_{\mu} q \). The interactions of constituent quarks with the scalar and vector mesons are described by the second term \(\mathcal{L}_{qm}\) \cite{Wang:2001jw} and can be expressed as
\begin{align}
\mathcal{L}_{qm} &= g_s \left( \bar{\Psi}_{L} M \Psi_{R} + \bar{\Psi}_{R} M^{\dagger} \Psi_{L} \right) - g_v \left( \bar{\Psi}_{L} \gamma^{\mu} l_\mu \Psi_{L} + \bar{\Psi}_{R} \gamma^{\mu} r_\mu \Psi_{R} \right) \nonumber \\
&= \frac{g_s}{\sqrt{2}} \bar{\Psi} \left[ \sum_{a=0}^8 s_a \lambda_a + i \gamma^5 \sum_{a=0}^8 p_a \lambda_a \right] \Psi - \frac{g_v}{2 \sqrt{2}} \bar{\Psi} \left[ \gamma^{\mu} \sum_{a=0}^8 v_{\mu}^a \lambda_a - \gamma^{\mu} \gamma^5 \sum_{a=0}^8 a_{\mu}^a \lambda_a \right] \Psi, \qquad 
\end{align}
where \(\Psi = (u,d,s)\) denotes the quark field corresponding to three flavors. The parameters \(g_s\) and \(g_v\) in the above equation are associated with couplings of quarks with scalar and vector meson fields. The chiral invariant term \(\mathcal{L}_{\Sigma \Sigma}\) and \(\mathcal{L}_{VV}\) of Eq. \eqref{L} are self-interaction terms for scalar and vector mesons, respectively \cite{Wang:2001jw}. For scalar mesons, the self-interaction term can be given as
\begin{align}
\mathcal{L}_{\Sigma \Sigma} &= -\frac{1}{2}k_0 \chi^2 \left(\sigma^2 + \zeta^2 + \delta^2\right) + k_1 \left(\sigma^2 + \xi^2 + 8^2\right)^2 + k_2 \left(\frac{\sigma^4}{2} + \frac{\delta^4}{2} + 3 \sigma^2 \delta^2 + \zeta^4 \right) \\
&+ k_3 \chi \left(\sigma^2 - \delta^2\right) \zeta - k_4 \chi^4 
- \frac{1}{4} \chi^4 \ln \frac{\chi^4}{\chi_0^4} + \frac{\xi}{3} \chi^4 \ln \left(\left(\frac{(\sigma^2 - \delta^2)\zeta}{\sigma_0^2 \zeta_0^2}\right)  \right) \left(\frac{\chi^3}{\chi_0^3}\right). \label{24}
\end{align}
For the vector mesons, we have
\begin{align}
\mathcal{L}_{VV} &= \frac{1}{2} \frac{\chi^2}{\chi_0^2} \left(m_\omega^2 \omega^2 + m_\rho^2 \rho^2 \right)+ g_4 \left(\omega^4 + 6 \omega^2 \rho^2 + \rho^4 \right).
\end{align}
The trace of the energy-momentum tensor is proportional to the fourth power of the dilaton field \(\chi\) because the last three terms in Eq. \eqref{24} are incorporated into the model to account for the trace anomaly in QCD.
 The QCD \(\beta\)-function at the one-loop level for three colors and three flavors is usually used to calculate the order of magnitude for the parameter \(\xi\) employed in the calculations.  The masses of the \(\pi\) meson (\(m_\pi\)), \(K\) meson (\(m_K\)), and the average mass of the \(\eta\) and \(\eta'\) mesons are utilized to compute the parameters \(k_0, k_1, k_2, k_3\), and \(k_4\), which appear in Eq.~\eqref{24}.\par
 Using the relations, \(\sigma_0 = -f_\pi\) and \( \zeta_0 = \frac{1}{\sqrt{2}}(f_\pi - 2f_K)\), the vacuum expectation values of the scalar meson fields, represented as \(\sigma\) and \(\zeta\) (i.e., \(\sigma_0\) and \(\zeta_0\)), can be expressed in terms of the pion decay constant (\(f_\pi\)) and the kaon decay constant (\(f_K\)). The fifth term in Eq.~\eqref{L}, the Lagrangian density \(\mathcal{L}_{\chi SB}\), corresponds to the explicit symmetry breaking term and is given by \cite{Wang:2001jw}
\begin{equation}
    L_{\chi SB} =  \frac{\chi^2}{\chi_0^2} \left[ m_\pi^2 f_\pi \sigma + \bigg( \sqrt{2}m_K^2 f_K - \frac{m_\pi^2}{\sqrt{2}} f_\pi \bigg)\zeta \right].
\end{equation}
There is one more explicit symmetry breaking term in Eq.~\eqref{L} (second last term), which is given by
\begin{align}
    L_{\triangle m} = -(\triangle m)\bar{\psi} S_1 \psi,
\end{align}
where \(S_1\) is the strange quark matrix operator and is defined by \( S_1 = \frac{1}{\sqrt{3}} \left( I - \frac{\lambda_8}{\sqrt{3}} \right) = \text{diag}(0, 0, 1)
\).
The last term of Eq. \eqref{L}, which represents the confinement of quarks within baryons, is expressed as
\begin{align}
    L_{c} = -\bar{\psi} \chi_c \psi,
\end{align}
with \(\chi_c\) being the scalar-vector potential, given by \(\chi_c (r) = \frac{1}{4}k_c r^2 (1 + \gamma_0)\). The value of coupling constant \(k_c\) is taken to be 98MeV .fm\(^{-2}\) \cite{Wang:2001jw}. \par
We have employed the mean-field approximation to study the properties of isospin asymmetric nuclear medium at finite temperature and density. The Dirac equation for the quark field \(\Psi_{qi}\) in the presence of the meson mean field is given by \cite{Wang:2001jw}
\begin{align}
    \bigg[- i \boldsymbol{\alpha} \cdot \boldsymbol{\nabla} + \chi_c(r) + \beta m^*_q \bigg] \Psi_{qi} = e^*_q \Psi_{qi}.
\end{align}
Here, the subscripts $q$ denotes the quark type $q = u, d, s$ within a baryon of type $i = p, n$. The effective quark mass $m_q^*$ and energy $e_q^*$ can be expressed through the following relations, which involve scalar fields ($\sigma, \zeta, \delta$) and vector fields ($\omega, \rho$), respectively
\begin{align}
    m^{\ast}_q = - g^{q}_{\sigma} \sigma - g^{q}_{\zeta} \zeta - g^{q}_{\delta} \mathbf{\mathit{I}}^{3q} \delta + m_{q0}, \label{34}
\end{align}
and
\begin{align}
    e^*_q = e_q - g^q_{\omega} \omega - g^q_{\rho} \mathbf{\mathit{I}}^{3q} \rho.
\end{align}
For quarks interacting with the scalar fields $\sigma$, $\zeta$, and $\delta$, the corresponding coupling constants are $g^q_\sigma$, $g^{q}_{\zeta}$, and $g^{q}_{\delta}$, respectively. Additionally, $I^{3q}$ represents the third component of the isospin for each quark flavor, with $I^{3u} = -I^{3d} = \frac{1}{2}$ and $I^{3s} = 0$. The additional mass term in Eq. \eqref{34} is \( m_{q0}\) which is zero for \(u\) and \(d\) quarks and 77 MeV for \(s\) quark. 
The effective mass of the $i$-th baryon is related to its effective energy $E_i^*$ and the in-medium spurious center-of-mass momentum $p_{i \text{ cm}}^{*}$ as
\begin{align}
    \textbf{\textit{M}}^*_i = \sqrt{ \bigg(\sum_q n_{qi} e^{*}_q + E_{i \text{ spin}} \bigg)^{2} - \langle p^{*2}_{i\text{ cm}} \rangle}.
\end{align}
Here, $n_{qi}$ represents the number of quarks of flavor $q$ in the $i$-th baryon. The term $E_{i\text { spin}}$ contributes as a correction factor to the baryon energy due to spin-spin interactions and is fitted to match the baryon vacuum mass. The spurious center-of-mass momentum $p_{i\text { cm}}^{*}$ can be written in terms of the in-medium constituent quark masses using the following relation \cite{Barik:1985rm, Barik:2013lna}
\begin{align}
    \langle p^{*2}_{i\text{ cm}} \rangle = \sum_q \frac{(11 e^*_q + m^*_q)}{6 (3 e^*_q + m^*_q) }(e^{*2}_q -m^{*2}_q).
\end{align}
To find the dependence of scalar and vector fields on baryonic density and temperature of the medium, we consider the thermodynamic potential for isospin asymmetric nuclear medium as 
\begin{align}
    \Omega = -\frac{k_B T}{(2\pi)^3} \sum_i \gamma_i \int_0^\infty d^3k \left\{ 
\ln\left(1 + e^{-[E_i^*(k) - \mu_i^*]/k_B T}\right) 
+ \ln\left(1 + e^{-[E_i^*(k) + \mu_i^*]/k_B T}\right)
\right\} - \mathcal{L}_M - V_{\text{vac}}. \label{38}
\end{align}
The summation runs over the nucleons in the medium, i.e., \( i = p, n\). The degeneracy factor for baryons is \( \gamma_i = 2 \), and the effective energy is given by \(
E_i^*(k) = \sqrt{M_i^{*2} + k^2}.
\)
The effective chemical potential \( \nu_i^* \) is related to the free chemical potential \( \nu_i \) by the following relation \cite{Wang:2001jw}
\begin{align}
\nu^*_i &= \nu_i - g^i_{\omega} \omega- g^i_ {\rho} \textbf{\textit{I}}^{3i} \rho.
\end{align}
Also, \( \mathcal{L}_M\) in Eq.~\eqref{38} can be expressed as \(\mathcal{L}_M=\mathcal{L}_{\Sigma \Sigma} +\mathcal{L}_{V V} + \mathcal{L}_{\chi SB}\). The thermodynamic potential given in Eq. \eqref{38} is minimized with respect to the scalar fields (\( \sigma, \zeta, \delta \)), the dilaton field (\( \chi \)), and the vector fields (\( \omega, \rho \)) in order to determine the baryonic density and temperature dependence of scalar and vector fields
\begin{align}
    \frac{\partial \Omega}{\partial \sigma} = \frac{\partial \Omega}{\partial \zeta} = \frac{\partial \Omega}{\partial \delta} = \frac{\partial \Omega}{\partial \chi} =\frac{\partial \Omega}{\partial \omega} =\frac{\partial \Omega}{\partial \rho} = 0. \label{40}
\end{align} 
For a given value of the isospin asymmetry parameter, \(\eta = -\frac{\Sigma_i \textbf{\textit{I}}_{3i} \rho_i}{\rho_B}\), the coupled equations for the scalar and vector fields are solved, where \(\rho_B\) is the medium's total baryonic density.

\section{Results and discussions} \label{sec3}
\subsection{In-medium effective meson masses} \label{sec3a}
In this section, we present an analysis of the medium-induced modifications to the masses, weak decay constants, and DAs of pseudoscalar and vector $B$ mesons, using the LFQM framework, which incorporates the medium-modified quark masses, determined by Eq. \eqref{34} within the CQMF model. The values of scalar fields $\sigma$, $\zeta \text{ and } \delta$, which depend on baryonic density ratio $\rho_B/\rho_{0}$ and temperature T, have been determined by solving the coupled system of equations, expressed in Eq. \eqref{40}. All the input parameters required to solve the system of equations of the scalar and vector fields within the CQMF model have been presented in Table \ref{tab:first}. The coupling constants of quarks with scalar and vector fields are related to each other by the following relations; $g^{s}_{\sigma} = g^{u}_{ \zeta}= g^{d}_{ \zeta}=0$, $\frac{g_s}{\sqrt{2}} = g^u_{\delta} = -g^d_{\delta} = g^u_{\sigma} = g^d_{\sigma} = \frac{1}{\sqrt{2}} g^s_{\zeta}$ and $g^u_{\omega} = g^d_{\omega} = \frac{g^N_{\omega}}{3}$.\par
As detailed in Ref. \cite{Choi:1997iq}, we compute the model parameters used in the LFQM by fitting mass spectra for the \(\eta_B\) and \(B^{\ast}\) mesons, which are listed in Table \ref{tab:second}. Using these parameters, the calculated mass and weak decay constant of the pseudoscalar (\(B^{+}, B^{0}\)) and vector (\(B^{\ast +}, B^{\ast 0}\)) mesons align well with the currently available experimental data \cite{ParticleDataGroup:2024cfk} , as outlined in Table \ref{tab:third}. 
\par

\begin{figure}[h]
    \centering
    \begin{subfigure}{0.48\textwidth}
        \centering
        \includegraphics[width=\textwidth]{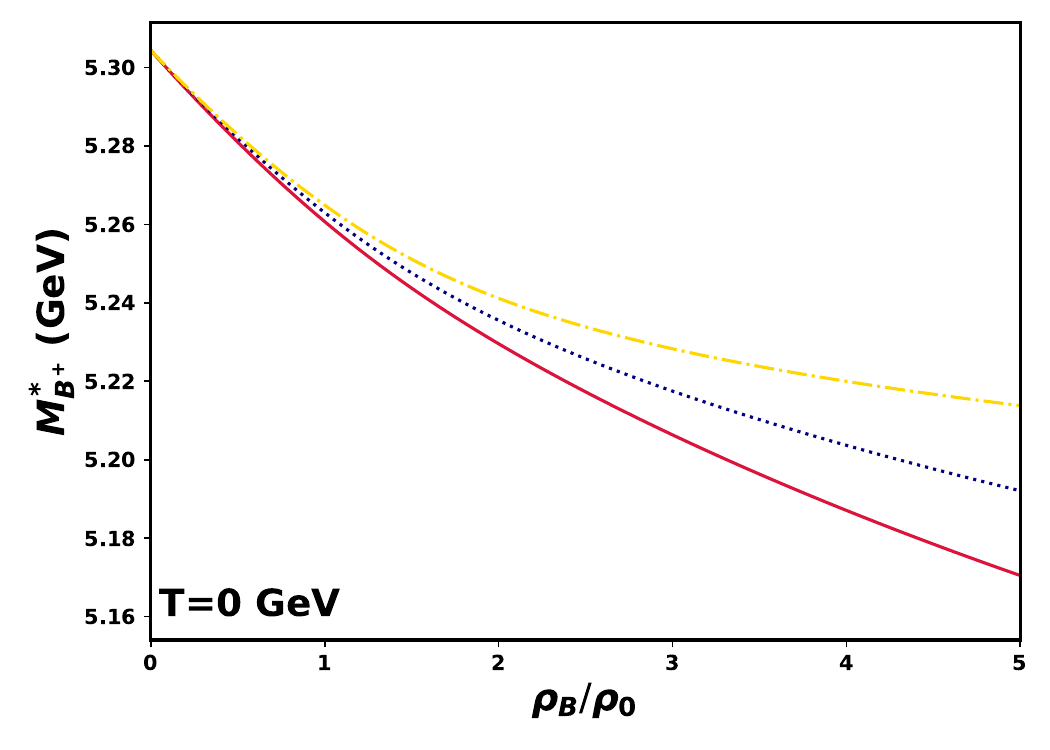}
    \end{subfigure}
    \hfill
    \begin{subfigure}{0.48\textwidth}
        \centering
        \includegraphics[width=\textwidth]{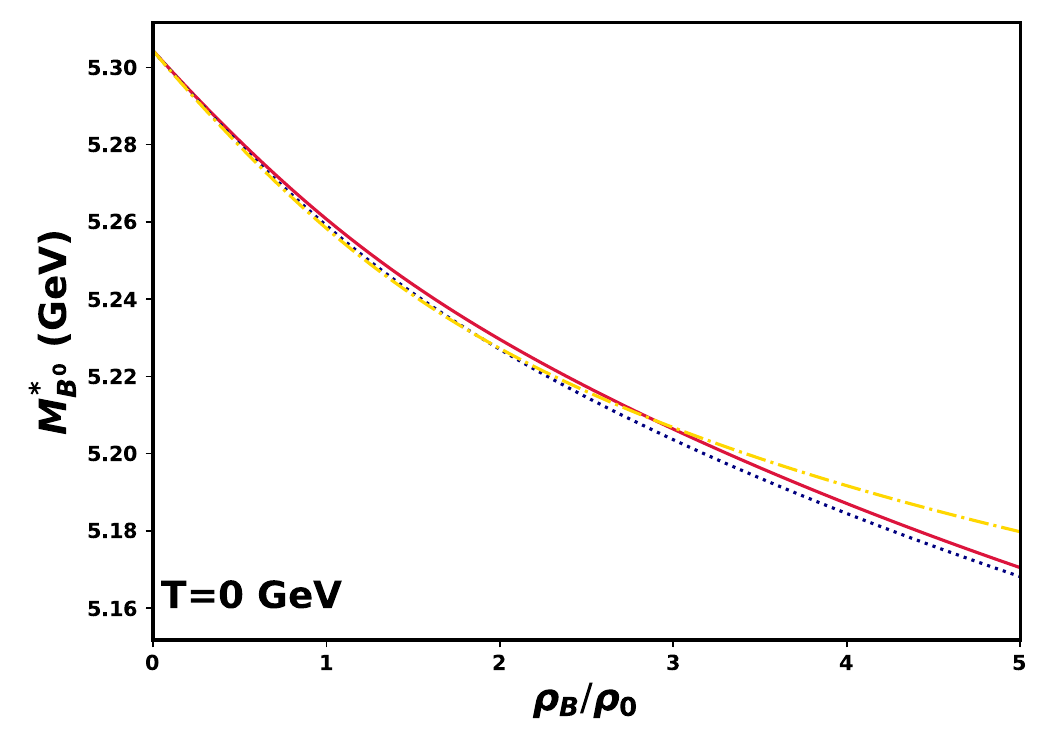}
    \end{subfigure}

    \begin{subfigure}{0.48\textwidth}
        \centering
        \includegraphics[width=\textwidth]{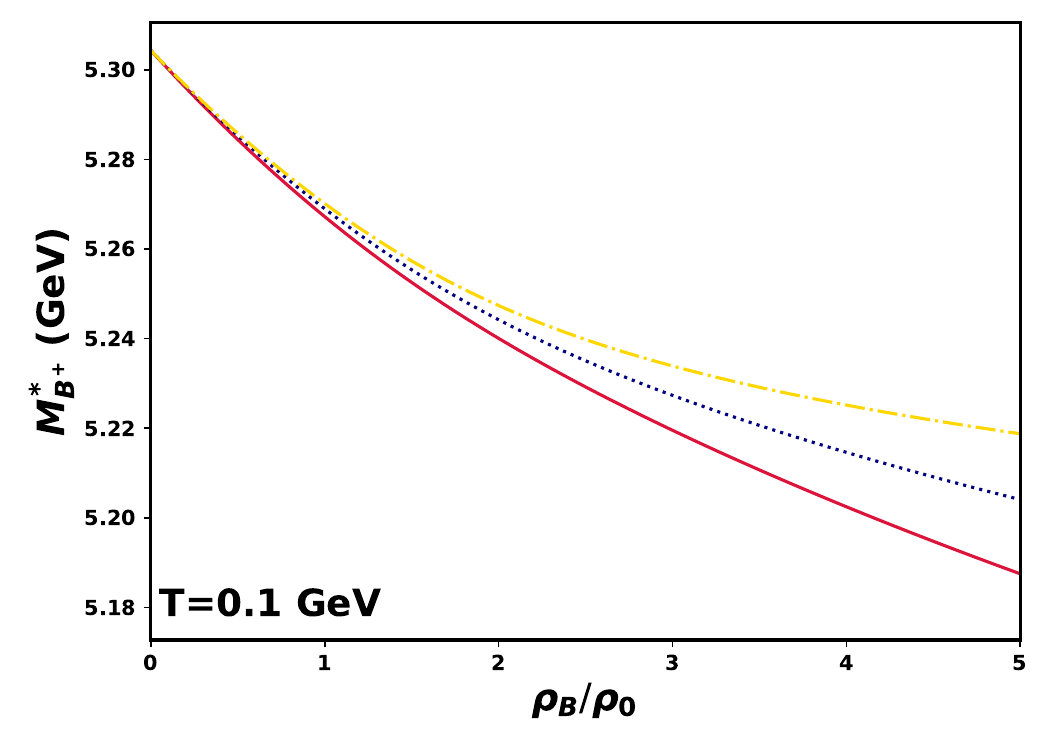}
    \end{subfigure}
    \hfill
    \begin{subfigure}{0.48\textwidth}
        \centering
        \includegraphics[width=\textwidth]{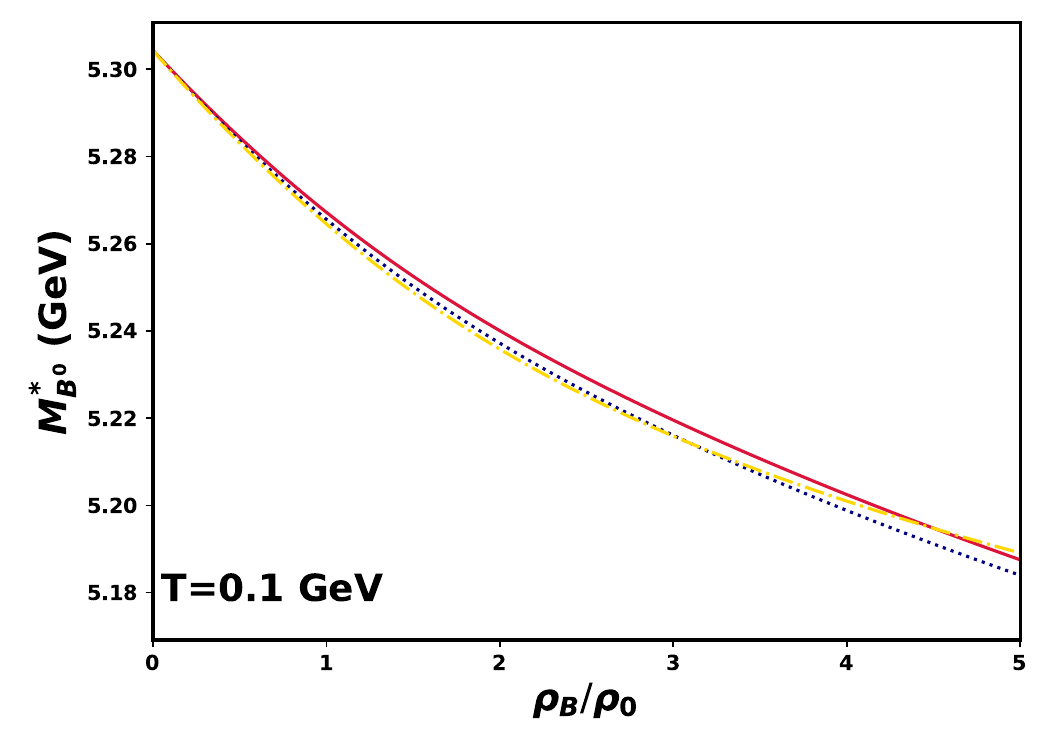}
    \end{subfigure}

    \begin{subfigure}{0.48\textwidth}
        \centering
        \includegraphics[width=\textwidth]{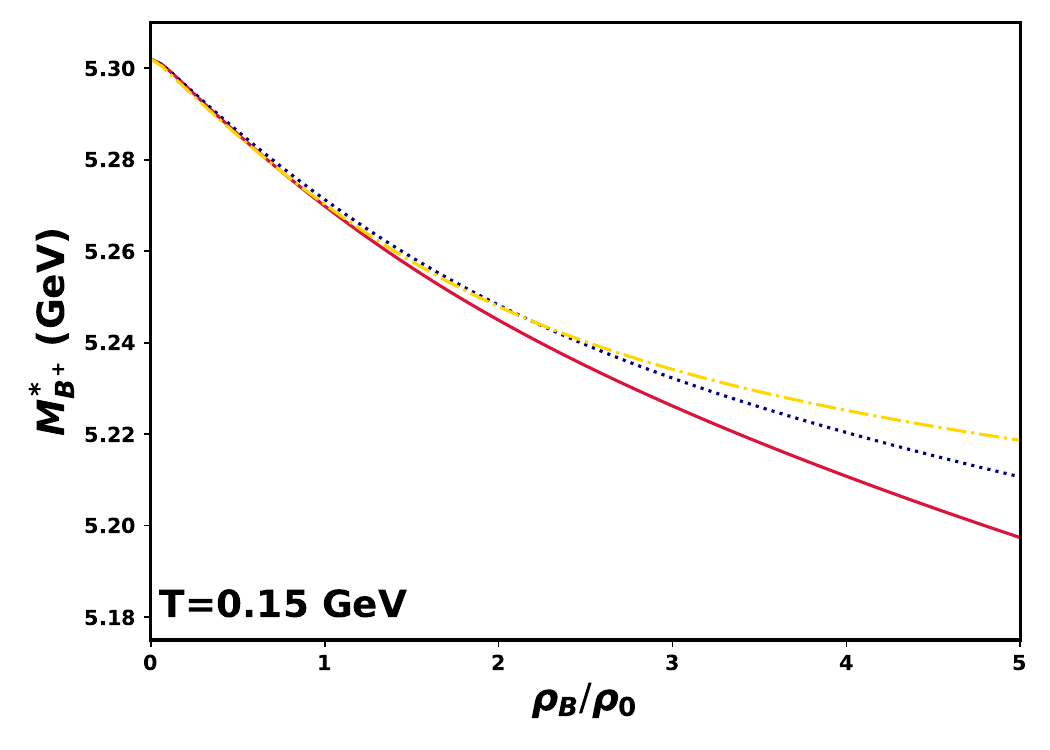}
    \end{subfigure}
    \hfill
    \begin{subfigure}{0.48\textwidth}
        \centering
        \includegraphics[width=\textwidth]{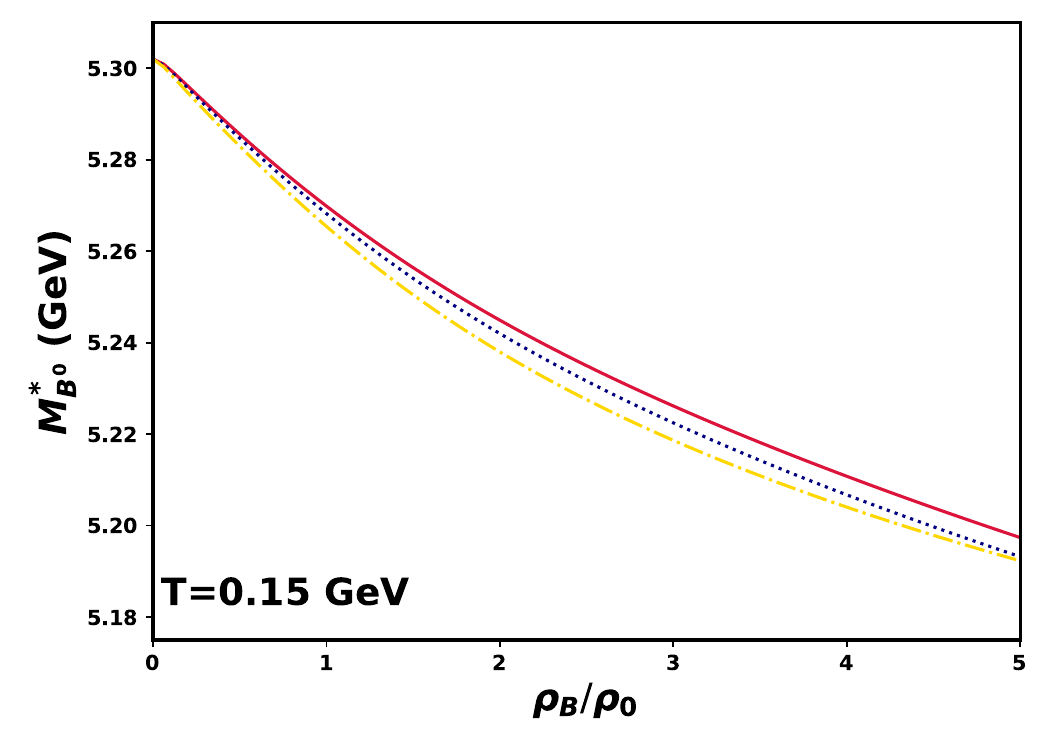}
    \end{subfigure}
\includegraphics[width=10 cm]{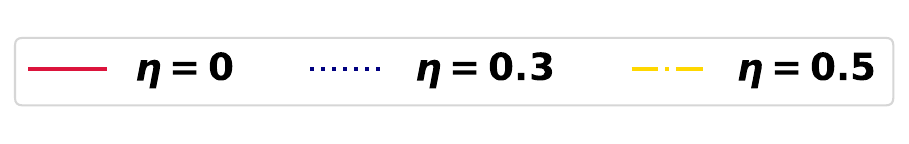}
    \caption{Effective masses of $B^{+}$ (left panel) and $B^{0}$ (right panel) mesons as a function of baryon density ratio \(\rho_B/\rho_0\) at temperatures T=0, 0.1 and 0.15 GeV and isospin asymmetry $\eta=0, 0.3  \text{ and }  0.5$.}
    \label{fig1}
\end{figure}

\begin{table}[h!]
\centering
\begin{tabular}{|c|c|c|c|c|c|c|c|c|c|}
\hline
\text{$k_0$} & \text{$k_1$} & \text{$k_2$} & \text{$k_3$} & \text{$k_4$} & \text{$\sigma_0$ [MeV]} & \text{$\zeta_0$ [MeV]} & \text{$\chi_0$ [MeV]} & \text{$\xi$} & \text{$\rho_0$ [fm$^{-3}$]} \\ \hline
4.94 & 2.12 & -10.16 & -5.38 & -0.06 & -92.8 & -96.5 & 254.6 & 6/33 & 0.16 \\ \hline
\textbf{$g^{u }_{\sigma} $} & \textbf{$ g^{u}_{ \zeta}$} & \textbf{$g^{u}_{\delta}$} & \textbf{$g^{s}_{ \zeta} = g_{s}$} & \textbf{$g_{4}$} & \textbf{$g^{p}_{ \delta}=g^{u}_{ \delta}$} & \textbf{$g^{N}_{ \omega}$} & \textbf{$g^{P}{\rho}$} & \text{$m_{\pi}$ [MeV]} & \text{$m_{K}$ [MeV]} \\ \hline
2.72 & 0 & 2.72 & 3.847 & 37.4 & 2.72 & 9.69 & 8.886 & 139 & 494 \\ \hline
\end{tabular}
\caption{Values of input parameters used in CQMF model for the present work \cite{Wang:2001jw}.}
\label{tab:first}
\end{table} 
\begin{table}[h]
    \centering
    \begin{tabular}{|c|c|c|c|c|c|c|c|}
        \hline
        \text{$m_q$} [GeV] & \textbf{$m_b$} [GeV] & \textbf{$\sigma'$} [GeV] & \text{a} [GeV] & \text{b} [GeV$^{2}$] & \textbf{$\alpha_s$} & \text{$\beta_{B^{+}(B^{0})}$} [GeV] & \text{$\beta_{B^{\ast+}(B^{\ast0})}$}[GeV] \\ 
        \hline
        0.2564 & 4.88 & 0.451 &-0.4050 & 0.18 & 0.3028 & 0.53633 & 0.51889\\ 
        \hline
    \end{tabular}
    \caption{Value of optimized quark masses (\(m_q, m_b\)), variational parameter \(\beta\), and potential parameters (\(\sigma', \text{a, b}, \alpha_{s}\)) obtained from the variational principle in the present work (where \(q = u \text{ and } d\)).}
    \label{tab:second}
\end{table}
\begin{table}[h]
    \centering
    \begin{tabular}{|c|c|c|c|c|c|}
        \hline
        \text{Mesons} & \text{$M_{theo}$ [GeV]} & \text{$M_{exp}$ [GeV]} & \text{$f_{theo}$ [GeV]} & \text{$f_{exp}$ [GeV]} \\ 
        \hline
       \text{$B^{+}(B^{0})$} & 5.304 & 5.279 &0.181 & 0.188  \\ 
        \hline
    \text{$B^{\ast+}(B^{\ast0})$}  & 5.324 & 5.325 &0.189 & .....   \\ 
        \hline
    \end{tabular}
    \caption{Comparison of masses and weak decay constants of pseudoscalar  (\(B^{+}, B^{0}\)) and vector (\(B^{\ast +}, B^{\ast 0}\)) mesons with experimental data \cite{ParticleDataGroup:2024cfk}.}
    \label{tab:third}
\end{table} 
We substitute the modified quark masses from Eq. \eqref{34} in Eq. \eqref{17} and get the expression of the effective mass of the mesons as
 \begin{equation}
     M^{\ast}_{\bar{q}q} = \frac{\beta}{\sqrt{\pi}} \sum_{i=q,\bar{q}} z^{\ast}_i e^{-z_i^{\ast2}/2} K_1\left(\frac{z^{\ast}_i}{2}\right) + a + \frac{2b}{\beta\sqrt{\pi}} - \frac{8\alpha_s ~\beta}{3\sqrt{\pi}} + \frac{32\alpha_s~ \beta^3 \langle \mathbf{S}_q \cdot \mathbf{S}_{\bar{q}} \rangle}{9\sqrt{\pi} m^{\ast}_q m_{\bar{q}}^{\ast}}, 
 \end{equation}
with modified \(z^{\ast}_{i}=m^{\ast2}_{i}/ \beta^2\). In Fig. \ref{fig1}, the effective masses of both pseudoscalar mesons $B^{+} (u\bar{b})$ and $B^{0} (d\bar{b})$ are plotted (in the left and right panels, respectively) as a function of the baryonic density ratio \(\rho_B/\rho_0\) for different values of isospin asymmetry $\eta$ and temperature T (GeV).
For the case of zero temperature (row 1, left panel), the dependence of the effective mass of $B^{+}$ meson is presented for different values of \(\eta=\) 0, 0.3 and 0.5. We have observed that for an isospin symmetric nuclear medium, i.e. \(\eta=0\), the effective mass of \(B^{+}\) meson falls off rapidly as a function of \(\rho_B/\rho_0\). The rate of declination in effective mass of \(B^{+}\) slows down as a function of the ratio \(\rho_B/\rho_0\) when the isospin asymmetry \(\eta\) is raised from 0 to 0.3 and 0.5. For the \(B^{0}\) meson (row 1, right panel), the effective mass behaves similarly to that of the \(B^{+}\) meson as a function of baryonic density ratio \(\rho_B/\rho_0\); it decreases sharply as the baryonic density increases. However, when the isospin asymmetry \(\eta\) is increased from 0 to 0.3, 
it has been observed that the rate of declination of the effective mass of \(B^{0}\) meson quickens as a function of baryonic density ratio \(\rho_B/\rho_0\). For an isospin asymmetry  \(\eta\) $=0.5$, the effective mass of $B^0$ meson falls off rapidly upto $\rho_B/\rho_0 \approx 2$ and then, slows down. This contradictory behavior of \(B^{+}\) and \(B^{0}\) mesons attributes to the mass splitting within the isospin quark doublet (\(u,d\)). This splitting is caused by the non-zero value of the \(\delta\) meson term, along with \(I^{3u}=\frac{1}{2}\) and \(I^{3d}=\frac{-1}{2}\), as shown in Eq. \eqref{34} in an isospin asymmetric medium. \\  
The impact of temperature on the effective masses of both pseudoscalar $B^{+}$ and $B^{0}$ mesons as a function of baryonic density ratio \(\rho_B/\rho_0\) can be observed by moving downward in the left and right panels, respectively. We have observed that for the \(B^{+}\) meson, as the temperature of the isospin symmetric and asymmetric medium increases from \(T = 0\) to \(0.1\) GeV, 
and then to \(0.15\) GeV, the rate of declination of effective mass decreases.
This phenomenon is similarly observed for the effective mass of the \(B^{0}\) meson as the temperature increases from T=0 to 0.1 and 0.15 GeV in an isospin symmetric and asymmetric nuclear medium. At non-zero temperature and $\eta=0.5$, the rapid fall of the effective meson mass is found to be consistent even at higher values of baryonic density of the medium, which is absent at zero temperature.\par
\begin{figure}[h]
    \centering
    \begin{subfigure}{0.48\textwidth}
        \centering
        \includegraphics[width=\textwidth]{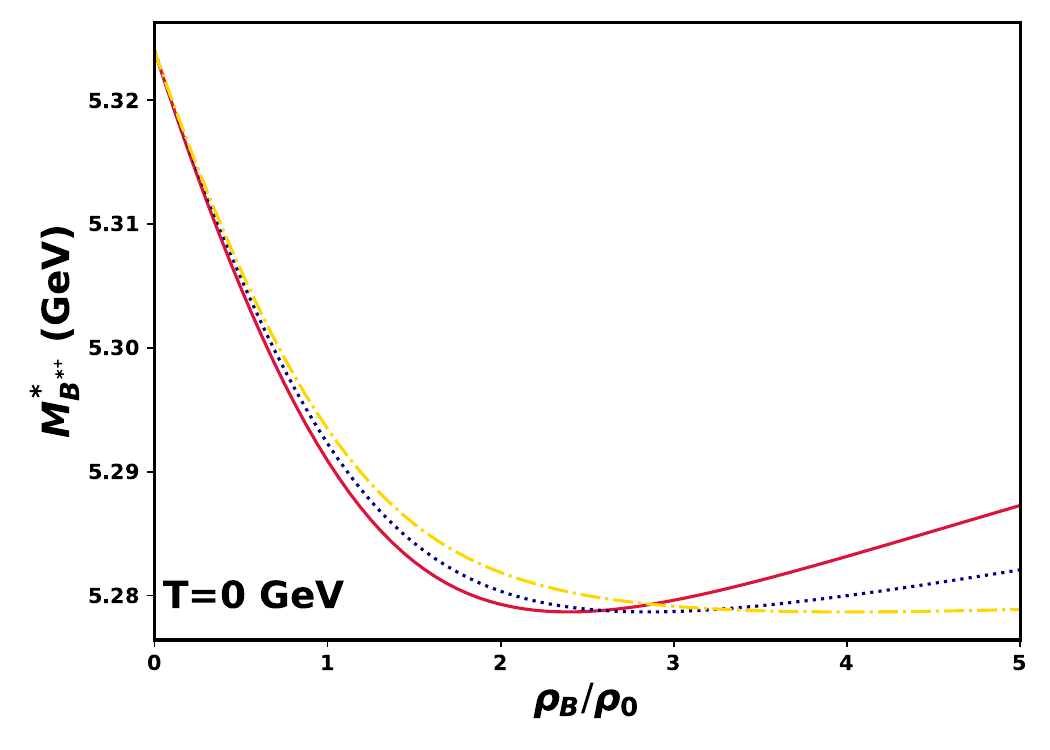}
    \end{subfigure}
    \hfill
    \begin{subfigure}{0.48\textwidth}
        \centering
        \includegraphics[width=\textwidth]{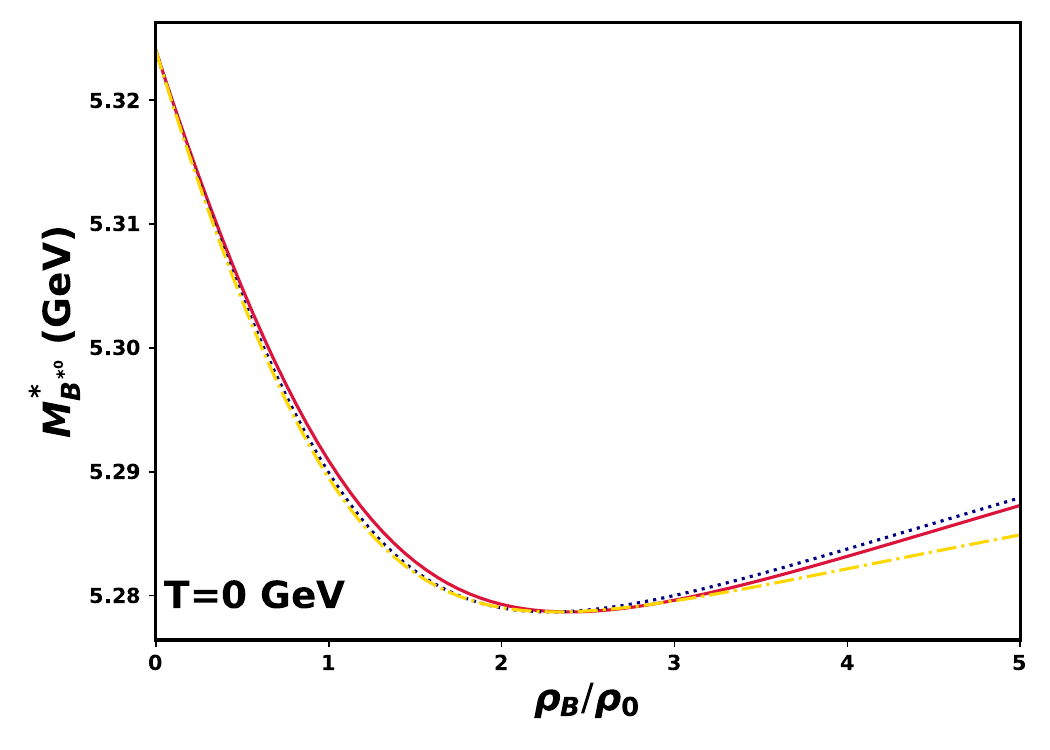}
    \end{subfigure}

    \begin{subfigure}{0.48\textwidth}
        \centering
        \includegraphics[width=\textwidth]{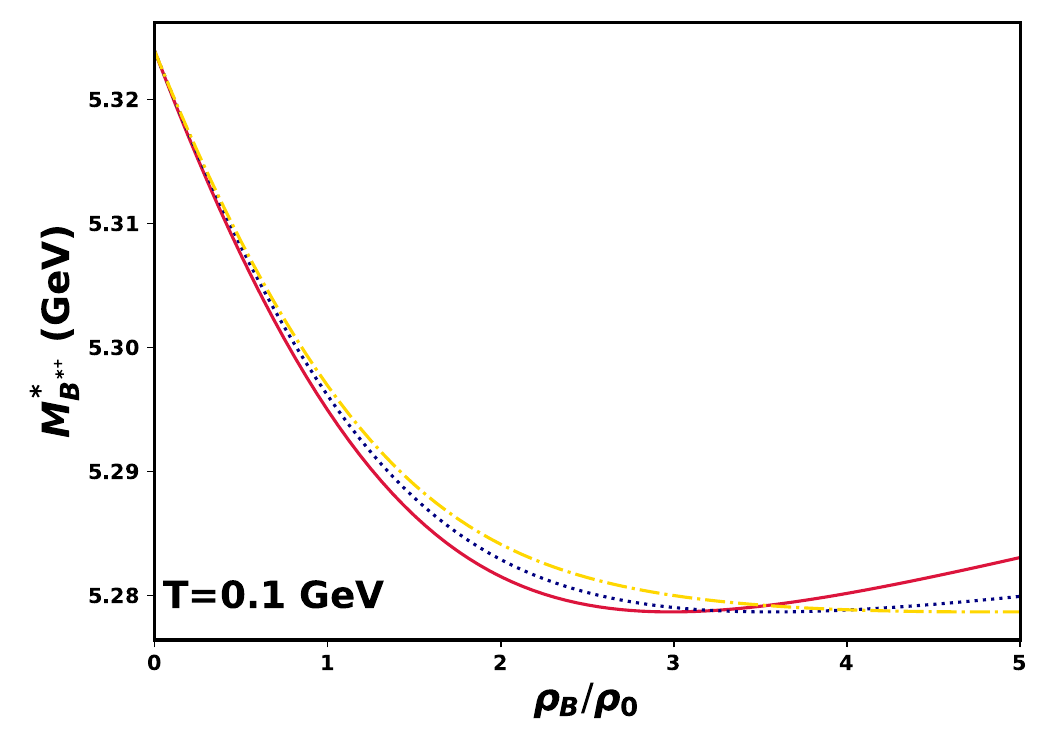}
    \end{subfigure}
    \hfill
    \begin{subfigure}{0.48\textwidth}
        \centering
        \includegraphics[width=\textwidth]{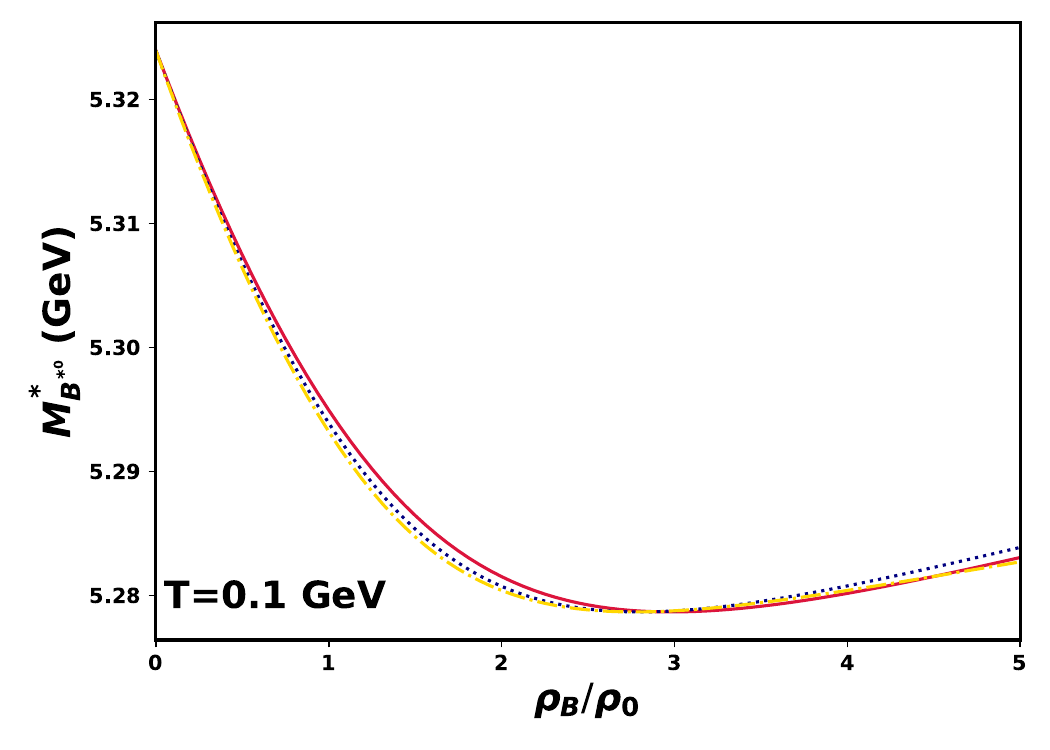}
    \end{subfigure}

    \begin{subfigure}{0.48\textwidth}
        \centering
        \includegraphics[width=\textwidth]{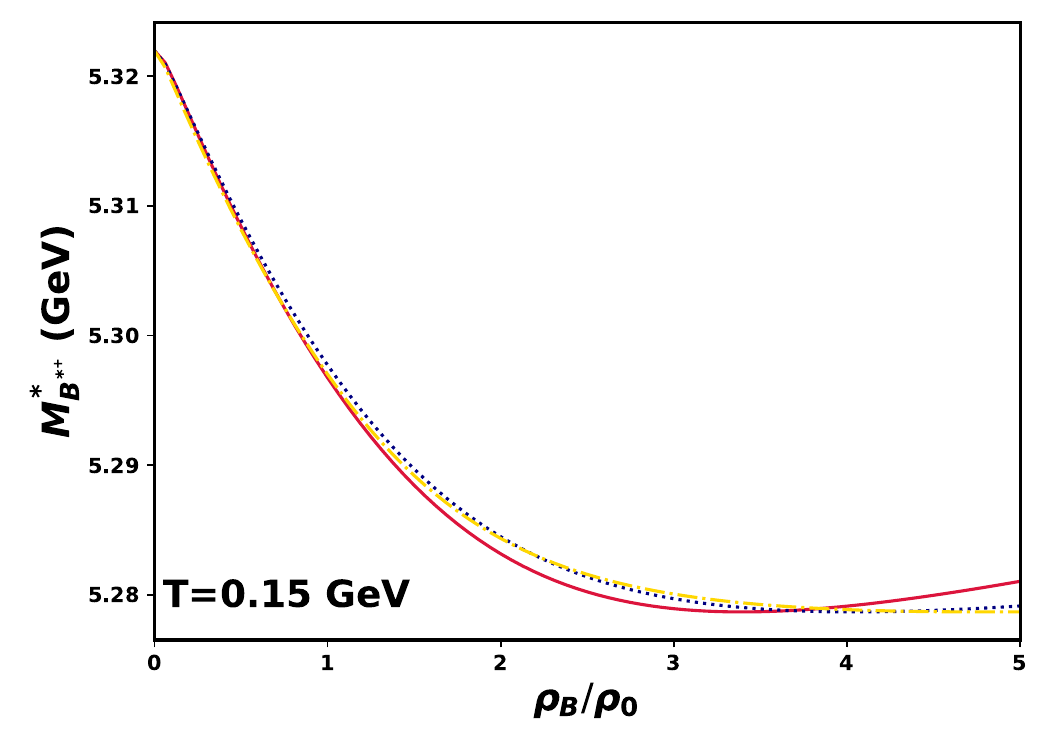}
    \end{subfigure}
    \hfill
    \begin{subfigure}{0.48\textwidth}
        \centering
        \includegraphics[width=\textwidth]{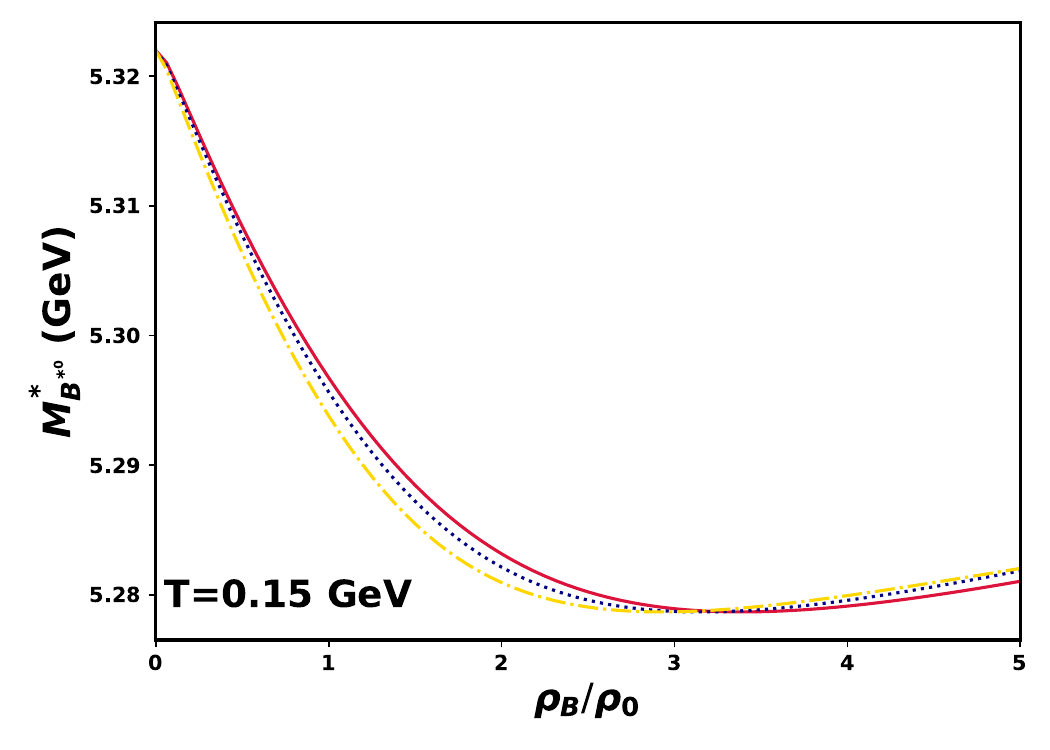}
    \end{subfigure}
\includegraphics[width=10 cm]{MASS/legends.pdf}
    \caption{Effective masses of $B^{\ast +}$ (left panel) and $B^{\ast 0}$ (right panel) mesons as a function of baryon density ratio \(\rho_B/\rho_0\) at temperatures T=0, 0.1 and 0.15 GeV and isospin asymmetry $\eta=0, 0.3 \text{ and } 0.5$.}
    \label{fig2}
\end{figure}

To study the effective masses of both vector mesons in an isospin asymmetric nuclear medium, we have plotted the variation of effective masses of  \(B^{\ast+} (u\bar{b})\) and \(B^{\ast0} (d\bar{b})\) meson (left and right panel, respectively) as a function of baryonic density ratio \(\rho_B/\rho_0\) for different values of asymmetry  \(\eta\) and temperature T (GeV) in Fig. \ref{fig2}. At \(T = 0\) GeV, the effective masses of the \(B^{\ast+}\) meson (row 1, left panel) are shown for discrete values of \(\eta=\) 0, 0.3, and 0.5. For isospin symmetric nuclear medium (\(\eta=0\)), we have observed that the effective mass of the \(B^{\ast +}\) meson falls rapidly up to \(\rho_B/\rho_0 \rightarrow2\), followed by a sudden increase. This unusual increase in the effective mass of \(B^{\ast+}\) as the baryonic density approaches 2 \(\rho_0\) could be due to the presence of spin-dependent term in the hyperfine interaction potential (as expressed in Eq. \eqref{hyperfine}, which causes hyperfine splitting for heavy quarks \cite{Choi:2015ywa}. As we increase \(\eta\) from 0 to 0.3 and then to 0.5, we have noticed that upto \(\rho_B/\rho_0 \approx3\) the rate at which the effective mass decreases slows down with the baryonic density ratio \(\rho_B/\rho_0\) in an isospin asymmetric nuclear medium. At higher baryonic densities 
increasing the isospin asymmetry 
results in a decrease in the rate of increment in effective masses of the \(B^{\ast+}\) meson. The variation of effective masses of the \(B^{\ast0}\) (row 1, right panel) meson with the baryonic density ratio \(\rho_B/\rho_0\) is similar to that of \(B^{\ast+}\) meson. Furthermore, as the value of \(\eta\) is raised from 0 to 0.3 and 0.5, the effect of isospin asymmetry on \(B^{\ast0}\) contrary to that of \(B^{\ast+}\) meson in isospin asymmetric medium. It has been observed that the rate of declination in effective mass of \(B^{\ast0}\) mesons up to \(\rho_B/\rho_0 \approx2.5\) decreases as the medium becomes more isospin asymmetric, at higher baryonic densities the value of inclination of effective masses increases for \(B^{\ast0}\) mesons as well. This conflicting behaviour of both the vector (\(B^{\ast+}\), \(B^{\ast0}\)) mesons in isospin asymmetric nuclear medium is due to mass splitting caused by the different lighter quark content in both the mesons.
\par By moving downwards in the left and right panels of Fig. \ref{fig2}, the effect of temperature on the respective effective masses of both vector mesons \(B^{\ast+}\) and \(B^{\ast0}\) as a function of baryonic density ratio \(\rho_B/\rho_0\) can be observed. We have noted that for \(B^{\ast+}\) meson in isospin asymmetric nuclear medium, as the temperature is raised from T=0 to 0.1 GeV, the rate of declination in the effective masses slows down, this slower pace becomes more appreciable when the value of T is raised to 0.15 GeV. The impact of raising the temperature from T=0 to 0.1 and then 0.15 GeV for \(B^{\ast0}\) meson in isospin asymmetric medium is similar to that of \(B^{\ast+}\) meson.
The influence of baryonic density, isospin asymmetry, and temperature on the medium modified \(B\) meson masses, as examined through the QCD sum rule approach through the modification of quark and gluon condensates in the medium \cite{Tsushima:2002cc, Dhale:2018plh, Yasui:2012rw, Hilger:2008jg, Wang:2015uya}
 , displays a similar trend to our current work, but is more pronounced. 
\subsection{In-medium weak decay constants}
Incorporating the effective masses of constituent light quarks from Eq. \eqref{34}, the expression of the weak decay constant in Eq. \eqref{19} will be modified as follows\\
For pseudoscalar mesons
\begin{equation}
    f^{\ast}_{P} = 2\sqrt{6} \int^{1}_{0} \mathrm{d}x \int \frac{\mathrm{d}^{2}\textbf{k}_\perp}{2(2 \pi)^3} \frac{\mathbf{\Phi^{\ast}}(x, \textbf{k}_\perp)}{\sqrt{[(1-x)m^{\ast}_{q}+x m^{\ast}_{\bar{q}}]^{2}+ \textbf{k}^{2}_\perp}}\bigg[(1-x)m^{\ast}_{q}+x m^{\ast}_{\bar{q}}\bigg],
\end{equation}
and for vector mesons,
\begin{equation}
    f^{\ast}_{V} = 2\sqrt{6} \int^{1}_{0} \mathrm{d}x \int \frac{\mathrm{d}^{2}\textbf{k}_\perp}{2(2 \pi)^3} \frac{\mathbf{\Phi^{\ast}}(x, \textbf{k}_\perp)}{\sqrt{[(1-x)m^{\ast}_{q}+x m^{\ast}_{\bar{q}}]^{2}+ \textbf{k}^{2}_\perp}}\bigg[[(1-x)m^{\ast}_{q}+x m^{\ast}_{\bar{q}}]+\frac{2 \textbf{k}_\perp^2}{\textbf{\textit{M}}^{\ast}_{0} + m^{\ast}_{q} + m^{\ast}_{\bar{q}}}\bigg].
\end{equation}
Here, the medium modified ground state wave function is denoted by \(\mathbf{\Phi^{\ast}}(x, \textbf{k}_\perp)\). The medium modified longitudinal momentum fraction \(x^{\ast}\) has been discussed in Refs. \cite{Puhan:2024xdq, Singh:2024lra}. However, to simplify our calculations, we have constrained it to $x$ only in the present work. 

The ratio of in-medium weak decay constant to its free space $f_P^\ast/f_P$ for pseudoscalar ($B^{+}, B^{0}$) mesons, in an isospin asymmetric nuclear medium has been compared in Table \ref{tab:fourth} for the different values of \(\rho_B= \rho_0, 2\rho_0, 3\rho_0\) at three different values of isospin asymmetry \(\eta=0, 0.3, 0.5\) and temperature T=0, 0.1, 0.15 GeV. It can be noted that the values of weak decay constant ratio show a decrement from unity as the baryonic density increases. A similar kind of behavior is found in \cite{Arifi:2023jfe} where the dependence of baryonic density on weak decay constants is studied up to \(\rho_B=\rho_0\). However, the decrease is found to be more appreciable at lower values of baryonic density, for both \(B^{+}\) and \(B^{0}\) mesons. In the case of both pseudoscalar mesons \(B^{+}\) and \(B^{0}\), it can be analyzed that as the temperature of the medium increases from 0 to 0.1 GeV and then to 0.15 GeV at any finite value of \(\eta\), the weak decay constant ratios \(f_{B^{+}}^{\ast}/f_{B^{+}}\) and \(f_{B^{0}}^{\ast}/f_{B^{0}}\) decline at a slower rate. However, the effect of isospin asymmetry is observed to be different for the decay constant ratios of \(B^{+}\) and \(B^{0}\) mesons. The \(f_{B^{+}}^{\ast}/f_{B^{+}}\) ratio declines more slowly with increasing \(\eta\) at finite temperatures, showing a more significant effect at higher densities (\(\rho_B/\rho_0 \rightarrow 3\)). Although, the rate of declinition of the \(f_{B^{0}}^{\ast}/f_{B^{0}}\) ratio occurs at a faster pace at higher densities. This distinct behavior of weak decay constant ratios arises from the splitting between the effective masses of the lighter isospin doublet quarks (\(u,d\)) of \(B^{+}\) and \(B^{0}\) mesons at higher densities and isospin asymmetry. Moreover, it has been noted that the $d$ quark experiences less influence than the $u$ quark as the isospin asymmetry of the medium rises; this occurs because the greater effective mass of the $d$ quark, leading to its interactions with the medium being slightly weaker than those of the $u$ quark.\\
\begin{table}[h!]
\centering
\renewcommand{\arraystretch}{1.2}
\resizebox{\textwidth}{!}{
\begin{tabular}{|c|c|c|c|c|c|c|c|c|c|c|}
\hline
& & \multicolumn{9}{c|}{Medium modified decay constant ratios $f_P^\ast/f_P$}\\
\cline{3-11}
\raisebox{-2.5ex}{\shortstack{Pseudoscalar\\mesons}} & \multirow{3}{*}{\raisebox{3.8ex}{\(\rho_B/\rho_0\)}}  & \multicolumn{3}{c|}{T = 0 GeV} & \multicolumn{3}{c|}{T = 0.1 GeV} & \multicolumn{3}{c|}{T = 0.15 GeV} \\
\cline{3-11}
& & \(\eta=0\) &\(\eta=0.3\)& \(\eta=0.5\)  & \(\eta=0\) &\(\eta=0.3\)& \(\eta=0.5\) & \(\eta=0\) &\(\eta=0.3\)& \(\eta=0.5\) \\
\cline{1-11}
&1 &0.954 &0.957 &0.960 &0.962 &0.964 &0.965 &0.967 &0.968 &0.967 \\
\cline{2-11}
$B^{+}$& 2&0.916 &0.923 &0.930 &0.928 &0.933 &0.937 &0.964 &0.941 &0.940 \\
\cline{2-11}
& 3& 0.896 &0.905 &0.915 &0.907 &0.914 &0.921 &0.915 &0.921 &0.923 \\
\hline
&1 &0.954 &0.952 &0.951 &0.962 &0.960 &0.958 &0.967 &0.965 &0.961 \\
\cline{2-11}
$B^{0}$& 2&0.916 &0.913 &0.914 &0.928 &0.925 &0.923 &0.967 &0.933 &0.928 \\
\cline{2-11}
& 3& 0.896&0.894 &0.897 &0.907 &0.904 &0.903 &0.915 &0.911 &0.908 \\
\hline
\end{tabular}
}
\caption{Medium modified weak decay constant ratios $f_P^\ast/f_P$ for the pseudoscalar $B$ mesons at different values of baryonic density ratio \(\rho_B/\rho_0\), isospin asymmetry \(\eta\) and temperature T. }
\label{tab:fourth}
\end{table}
\begin{table}[h!]
\centering
\renewcommand{\arraystretch}{1.2}
\resizebox{\textwidth}{!}{
\begin{tabular}{|c|c|c|c|c|c|c|c|c|c|c|}
\hline
& & \multicolumn{9}{c|}{Medium modified decay constant ratios $f_V^\ast/f_V$}\\
\cline{3-11}
\raisebox{-2.5ex}{\shortstack{Vector\\mesons}} & \multirow{3}{*}{\raisebox{3.8ex}{\(\rho_B/\rho_0\)}}  & \multicolumn{3}{c|}{T = 0 GeV} & \multicolumn{3}{c|}{T = 0.1 GeV} & \multicolumn{3}{c|}{T = 0.15 GeV} \\
\cline{3-11}
& & \(\eta=0\) &\(\eta=0.3\)& \(\eta=0.5\)  & \(\eta=0\) &\(\eta=0.3\)& \(\eta=0.5\) & \(\eta=0\) &\(\eta=0.3\)& \(\eta=0.5\) \\
\cline{1-11}
&1 &0.963 &0.965 &0.969 &0.9690 &0.971&0.972 &0.973 &0.975 &0.973\\
\cline{2-11}
$B^{\ast+}$& 2&0.932 &0.937 &0.943 &0.942 &0.946 &0.949 &0.948 &0.951 &0.952 \\
\cline{2-11}
& 3& 0.914 &0.922 &0.931 &0.923 &0.930 &0.936 &0.930 &0.936 &0.937\\
\hline
&1 &0.963 &0.961 &0.960 &0.969 &0.967 &0.966 &0.973 &0.972 &0.969 \\
\cline{2-11}
$B^{\ast0}$& 2&0.932 &0.929 &0.9292 &0.942 &0.939 &0.937 &0.948 &0.945 &0.941\\
\cline{2-11}
& 3& 0.914 &0.913 &0.912 &0.923 &0.921 &0.920 &0.930 &0.927 &0.924 \\
\hline
\end{tabular}
}
\caption{Medium modified weak decay constant ratio $f_V^\ast/f_V$ for vector $B$ mesons at different values of baryonic density ratio \(\rho_B/\rho_0\), isospin asymmetry \(\eta\) and temperature T. }
\label{tab:fifth}
\end{table}
A quantitative analysis has also been provided for the weak decay constant ratios $f_V^\ast/f_V$ of the vector mesons in Table \ref{tab:fifth}. The weak decay constant ratios of vector mesons are also found to be decreased with the increment of baryonic density. It has been noted that the behavior of the ratio \(f_{B^{\ast+}}^{\ast}/f_{B^{\ast+}}\) resembles that of \(f_{B^{+}}^{\ast}/f_{B^{+}}\), and the ratio \(f_{B^{\ast0}}^{\ast}/f_{B^{\ast0}}\) shows similarities to \(f_{B^{0}}^{\ast}/f_{B^{0}}\) when considering variations in the values of the isospin asymmetry \(\eta\) and the temperature T (GeV). Furthermore, the variations in weak decay constant ratios among vector mesons in an isospin asymmetric nuclear medium are less pronounced than those observed in pseudoscalar mesons. This can be explained by the presence of the medium modified invariant meson mass \(\textbf{\textit{M}}^{\ast}_{0}\) in the denominator of Eq. \eqref{20}, which is utilized to calculate the weak decay constant of vector mesons \cite{Arifi:2023jfe}.\par
The analysis of the medium-modified weak decay constants of pseudoscalar and vector \(B\) mesons has been conducted using the QCD sum rule approach \cite{Wang:2015uya, Azizi:2014bba, Chhabra:2016vhp}. This study also explains how the weak decay constants are affected by increasing baryonic density, isospin asymmetry, and temperature, demonstrating consistency with our earlier work.

\subsection{In-medium distribution amplitudes}
The explicit expression of DA of mesons in Eq. \ref{23} modified in the medium is given by
\begin{align}
    \phi^{\ast}_{\mathrm{M}}(x) = \frac{2\sqrt{6}}{f^{\ast}_{\mathrm{M}}} \int \frac{\mathrm{d}^2 \mathbf{k}_{\perp}}{2(2\pi)^3} \frac{\mathbf{\Phi^{\ast}}(x, \mathbf{k}_{\perp})}{\sqrt{\mathcal{A}^{\ast2} + \mathbf{k}_{\perp}^2}} \mathcal{O}^{\ast}_{M}. 
\end{align}

\begin{figure*}
\centering
\begin{minipage}[c]{0.98\textwidth}
 \includegraphics[width=0.7\textwidth]{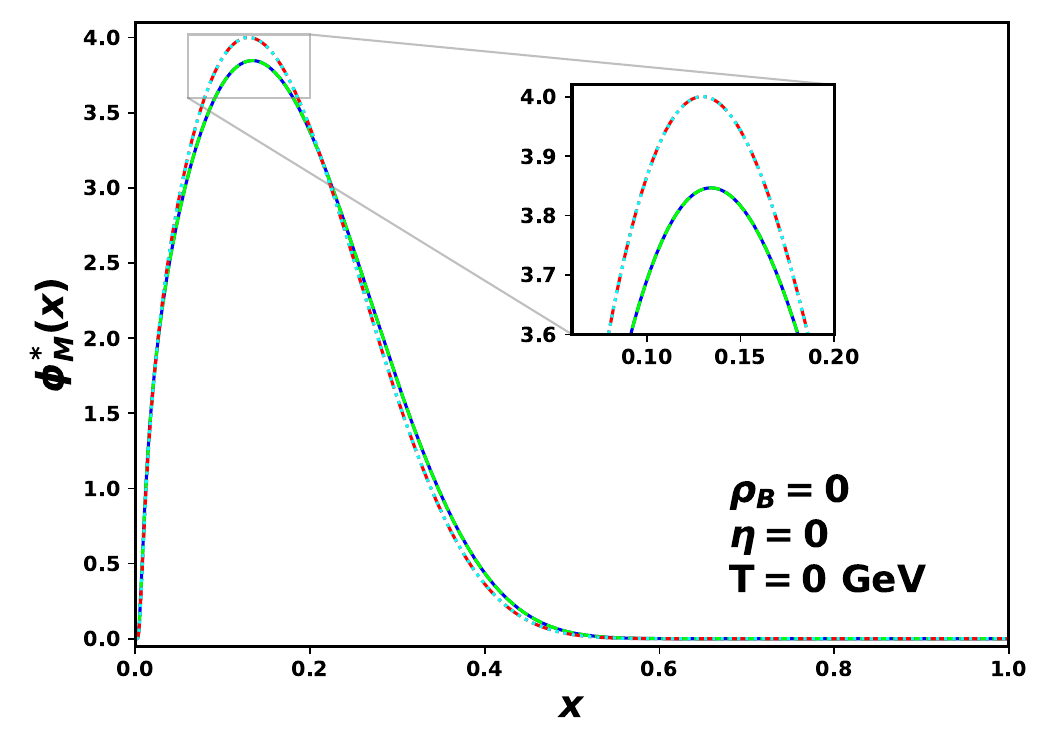}
 \hspace{3 cm}
\includegraphics[width=10 cm]{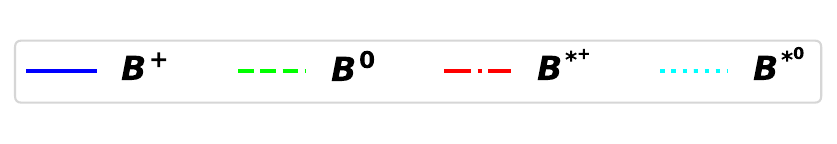}
\end{minipage}
\caption{\label{fig5} Distribution amplitudes as a function of longitudinal momentum fraction $x$ at baryonic density $\rho_B=0$, temperature T=0 GeV and isospin asymmetry $\eta=0$.}
\end{figure*}

In Fig. \ref{fig5}, we have plotted the variation of DAs \(\phi^{\ast}_{M} (x)\) for pseudoscalar ($B^{+}, B^{0}$) and vector ($B^{\ast+}, B^{\ast0}$) mesons as a function of longitudinal momentum fraction $x$, in free space, i.e., at zero baryonic density \(\rho_B\), isospin asymmetry \(\eta\), and temperature T (GeV). The lighter quark in the $B$ meson carries the longitudinal momentum fraction $x$, whereas the heavier quark carries $1-x$. The shape of the DAs for the $B$ mesons is asymmetric, exhibiting a sharp peak within the range of 0.1\(< x < \)  0.2, consistent with results obtained in Refs. \cite{Arifi:2023jfe, Arifi:2024mff}. It has been observed that, in free space, the peaks of pseudoscalar and vector $B$ mesons differ. This distinctive behavior primarily results from the mass differences among the constituent quarks. The peak associated with vector $B$ mesons lies higher in magnitude than that of pseudoscalar $B$ mesons, which is attributed to their greater mass and variational parameter \(\beta\). The peak values of \(\phi^{\ast}_{M} (x)\) for $B^{+}, B^{0}, B^{\ast+}, B^{\ast0}$ are found to be at $x=$ 0.133, 0.133, 0.129 and 0.129, respectively.
 
\begin{figure*}
\centering
\begin{minipage}[c]{1\textwidth}
(a)\includegraphics[width=7.0cm]{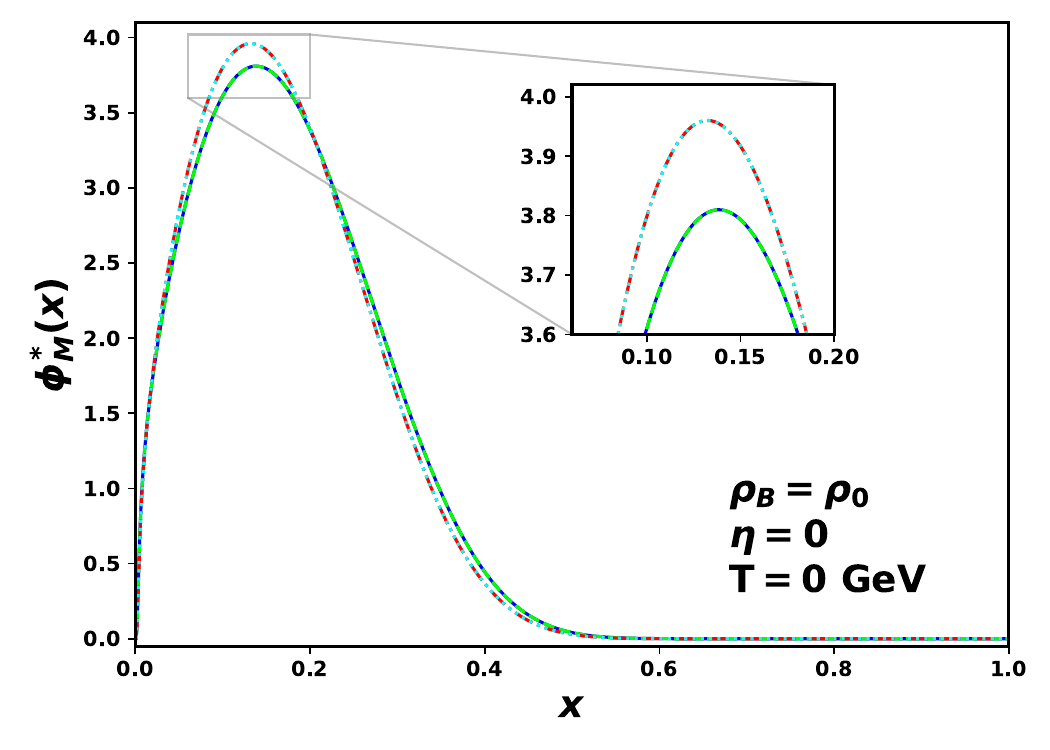}
\hspace{0.03cm}
(b)\includegraphics[width=7.0cm]{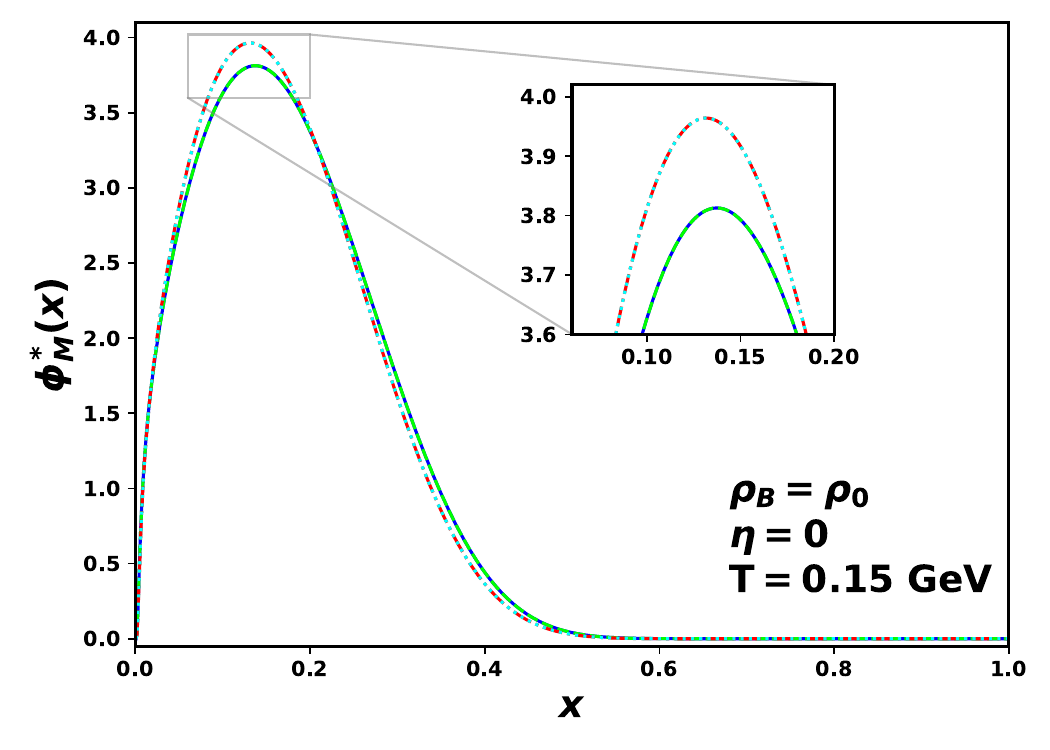}
\hspace{0.03cm}	\\
(c)\includegraphics[width=7.0cm]{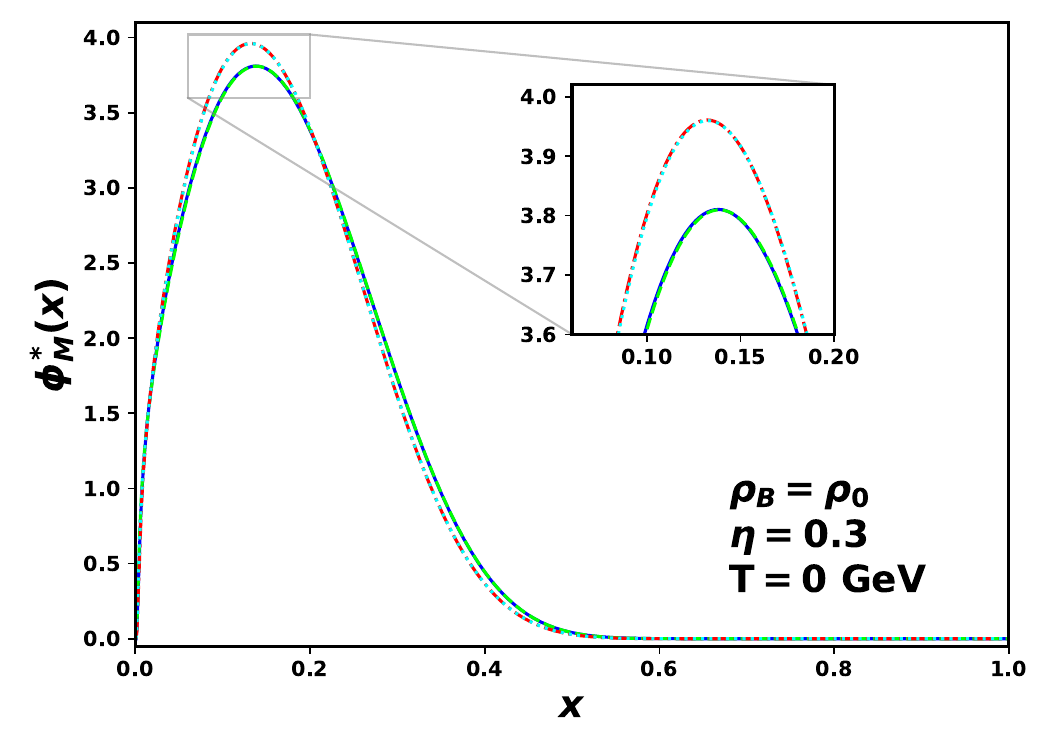}
\hspace{0.03cm}
(d)\includegraphics[width=7.0cm]{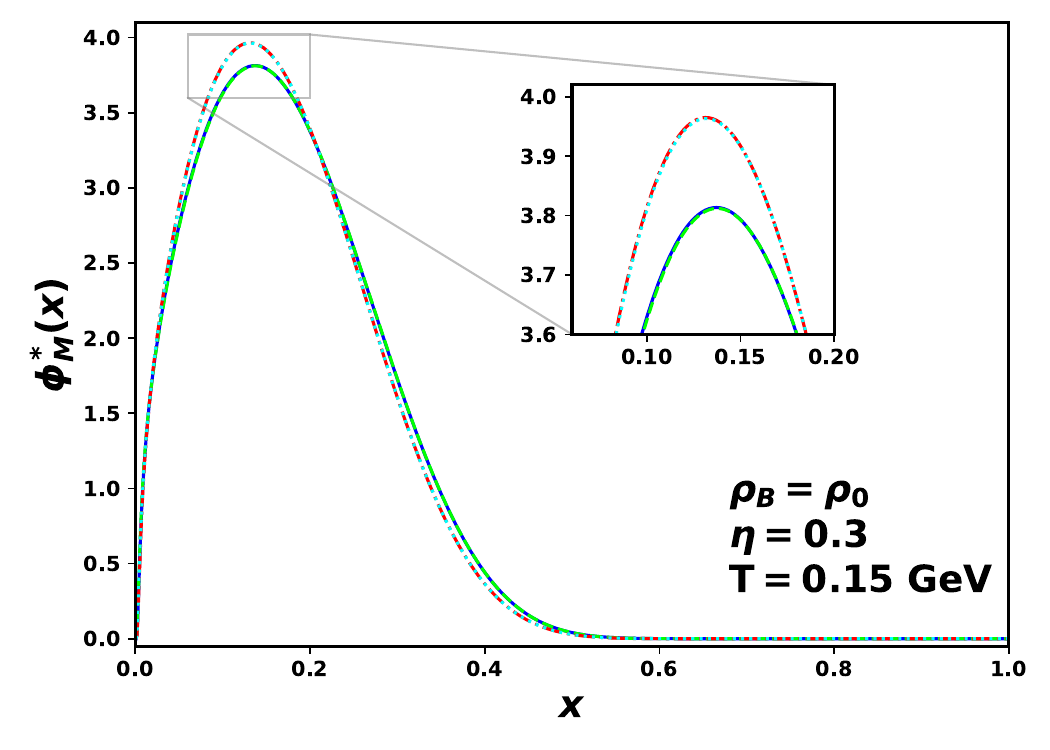}
\hspace{0.03cm}	\\
\includegraphics[width=10 cm]{DA/legendsDA.pdf}
\end{minipage}
\caption{\label{fig6} Distribution amplitudes as a function of longitudinal momentum fraction $x$ at baryonic density $\rho_B=\rho_0$. Results are shown for different isospin asymmetries ($\eta=0,0.3$) at temperature T=0 GeV (left panel) and T=0.15 GeV (right panel).}
\end{figure*}

In Fig. \ref{fig6}, we have displayed the plots of medium modified DAs \(\phi^{\ast}_{M}(x)\) for pseudoscalar ($B^{+}, B^{0}$) and vector ($B^{\ast+}, B^{\ast0}$) mesons as a function of \(x\) at nuclear saturation density \(\rho_B=\rho_0\) for isospin asymmetry \(\eta=\) 0 and 0.3, along with different temperatures T=0 and 0.15 GeV. The general trend of the \(\phi^{\ast}_{M}(x)\) is similar to that of free space. It has been observed that (in Fig. \ref{fig6}(a)) the peak values of the DAs of the \(B\) meson decrease as the density of the nuclear medium increases.
Now, the peak values for \(B^{+}, B^{0}, B^{\ast+}, B^{\ast0}\) are observed at \(x\) values of 0.136, 0.136, 0.132, 0.132, indicating a shift of peak on higher values of \(x\) with increasing baryonic density. This phenomenon is attributed to the reduction in effective quark masses \(m_q^{\ast}\) within the medium, a consequence of the partial restoration of chiral symmetry. As we move from the left panel to the right panel to study the impact of temperature, which is raised from 0 to 0.15 GeV in an isospin symmetric (row 1) and asymmetric (row 2) nuclear medium, we find a slight increment in the peaks of the DAs for all the mesons. In quantitative terms, the peak values of \(\phi^{\ast}_{M} (x)\) for $B^{+}, B^{0}, B^{\ast+}, B^{\ast0}$ were 3.860, 3.860, 3.960, 3.960, respectively in Fig. \ref{fig6}(a), and when the temperature is raised, the peak values become 3.863, 3.863, 3.964, 3.964, respectively in Fig. \ref{fig6}(b). A similar increase in the peak was seen in Refs. \cite{Puhan:2024xdq, Singh:2024lra} for pion and kaon as a function of temperature. In the case of pion, the difference was pronounced because it has a lighter $ q\bar{q}$ pair and both of them are influenced significantly by the nuclear medium. 
On moving from row 1 to row 2, we studied the impact of isospin asymmetry by increasing \(\eta\) from 0 to 0.3. We observed that the effect of the isospin asymmetry is less pronounced on the DA of the \(B\) mesons; however, a mass splitting occurs between mesons containing \(u\) quarks and those with \(d\) quarks. As illustrated in Fig. \ref{fig6}(c), the peak values for \((B^{+}, B^{\ast+})\) are found to be (3.861, 3.961), whereas for \((B^{0}, B^{\ast0})\), the values are (3.860, 3.960), respectively. Fig. \ref{fig6}(d) reveals a more pronounced splitting, with the peak values for \(B^{+}, B^{0}, B^{\ast+}, B^{\ast0}\) mesons are (3.863,3.862,3.865,3.963), respectively. \par
\begin{figure*}
\centering
\begin{minipage}[c]{1\textwidth}
(a)\includegraphics[width=7.0cm]{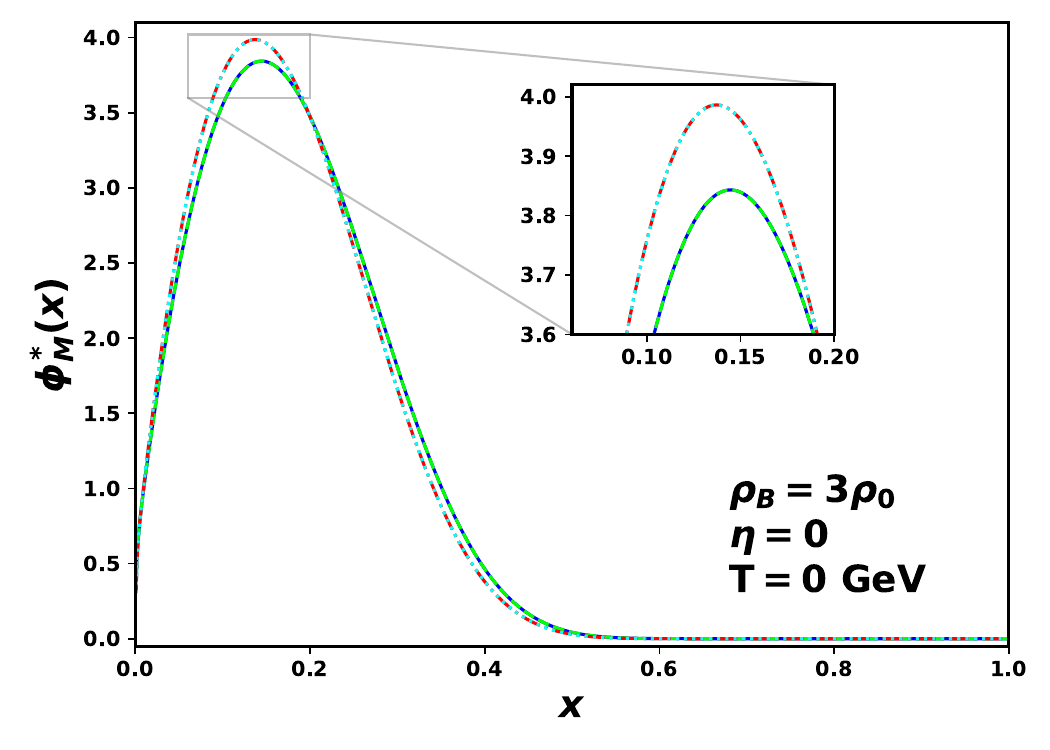}
\hspace{0.03cm}
(b)\includegraphics[width=7.0cm]{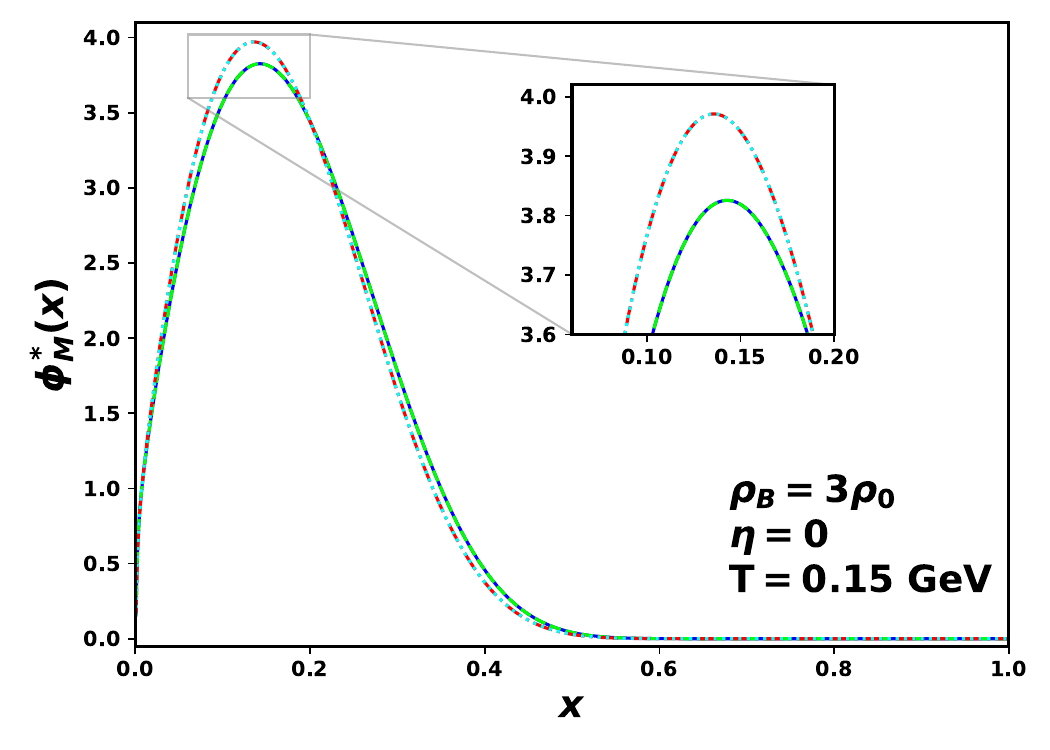}
\hspace{0.03cm}	\\
(c)\includegraphics[width=7.0cm]{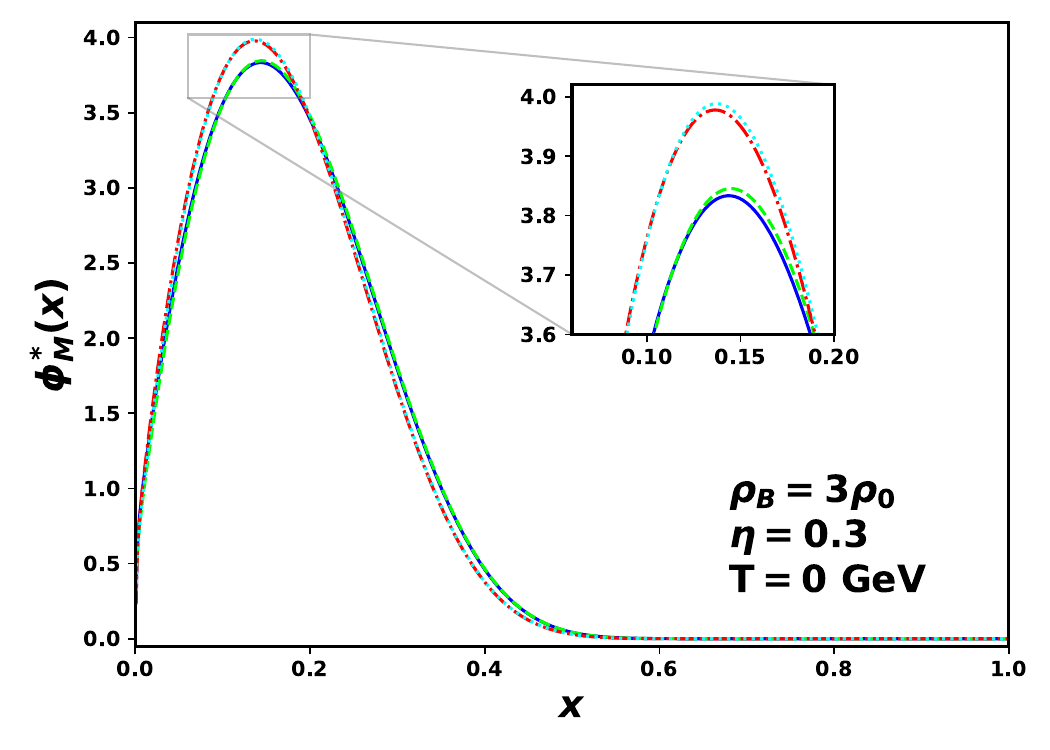}
\hspace{0.03cm}
(d)\includegraphics[width=7.0cm]{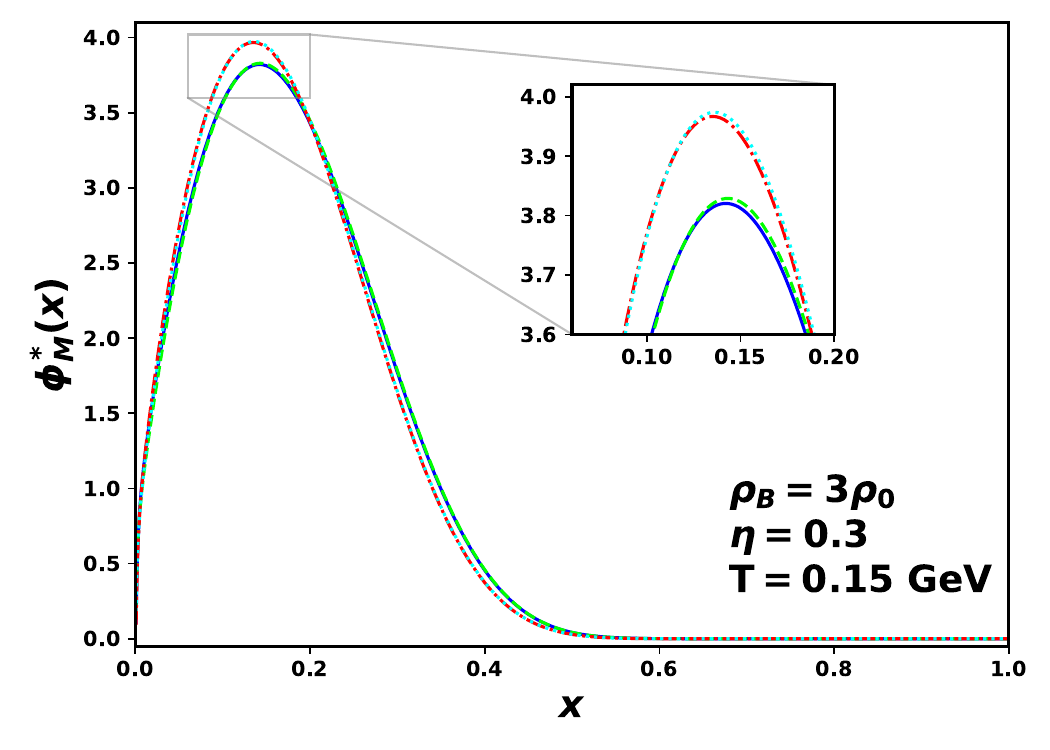}
\hspace{0.03cm}	\\
\includegraphics[width=10 cm]{DA/legendsDA.pdf}
\end{minipage}
\caption{\label{fig7} Distribution amplitudes as a function of $x$ at baryonic density $\rho_B=3 \rho_0$. Results are shown for different isospin asymmetries ($\eta=0,0.3$) at temperature T=0 GeV (left panel) and T=0.15 GeV (right panel).}
\end{figure*}

The plots of medium modified DAs as a function of $x$ have been shown in Fig. \ref{fig7} at baryonic density \(\rho_B=3\rho_0\) for different values of \(\eta\) as 0 and 0.3, and temperature T=0 and 0.15 GeV. We have observed a further shift in the peak values of \(\phi^{\ast}_{M} (x)\) towards greater \(x\) values in Fig. \ref{fig7}(a) at \(\rho_B=3\rho_0\), \(\eta=0\), and T=0 GeV, with the peak values for the $B^{+}, B^{0}, B^{\ast+}, B^{\ast0}$ mesons recorded at \(x=\) 0.140, 0.140, 0.136, 0.136, respectively. When the temperature of the nuclear medium is increased from T=0 (left panel) to 0.15 GeV (right panel), the peak values decrease from (3.886, 3.886, 3.986, 3.986) in Fig. \ref{fig7}(a) to (3.873, 3.873, 3.971, 3.971) in Fig. \ref{fig7}(b) for the ($B^{+}, B^{0}, B^{\ast+}, B^{\ast0}$) mesons, demonstrating a slower rate of effective mass reduction. This discrepancy in the peak values of DAs with increasing temperature, in comparison to pions and kaons mentioned in Refs. \cite{Puhan:2024xdq, Singh:2024lra},  arises from the predominance of the heavier \(b\) quark, as the masses of the lighter quarks diminish at higher baryonic densities. On increasing the isospin asymmetry of the nuclear medium from \(\eta=\) 0 (row 1) to 0.3 (row 2), we have observed the splitting between masses of \(B^{+}, B^{\ast+}\) and \(B^{0}, B^{\ast0}\) mesons. In Fig. \ref{fig7}(c), the peak values of \(B^{+}\) and \(B^{0}\) are observed to be 3.880 and 3.890 at \(x = 0.140\) and \(x = 0.141\), respectively. Additionally, the peak values for \(B^{\ast+}\) and \(B^{\ast0}\) are 3.978 and 3.988 at \(x = 0.136\) and \(x = 0.137\). The enhanced distinctions of effective masses is further illustrated in Fig. \ref{fig7}(d), where the peak values of the ($B^{+}, B^{0}, B^{\ast+}, B^{\ast0}$) mesons are (3.868, 3.876, 3.966, 3.974) at \(x\) (0.138, 0.139, 0.134, 0.135) respectively, signifying the influence of an isospin asymmetric nuclear medium at \(\rho_B/\rho_0=3\). Furthermore, it can be inferred from the aforementioned that, at $x>0.5$, the impact of medium on the DAs of pseudoscalar and vector $B$ mesons is essentially insignificant.

\section{Summary}\label{sec4}
In our present work, we have explored the in-medium effects of isospin asymmetric nuclear medium on masses, weak decay constants, and distribution amplitudes (DAs) of pseudoscalar ($B^{+}, B^{0}$) and vector ($B^{\ast +}, B^{\ast 0}$) mesons. To achieve this, we have computed the medium modified masses of quarks from chiral SU(3) quark mean field model (CQMF) and used them as input parameters in the light-front quark model (LFQM).
The free space mass eigenvalues of all the pseudoscalar and vector $B$ mesons have been computed. 
To optimize the model parameters, we have applied the variational principle, incorporating both the Coulomb and confining potentials, while treating the hyperfine interaction as a perturbation. 
We have analyzed the vacuum masses, weak decay constants, and DAs of pseudoscalar and vector $B$ mesons using the framework of  LFQM. Our calculated masses and weak decay constants for the 
$B$ mesons are in good agreement with experimental data. To incorporate the in-medium effects, we have used the CQMF model, which modifies the constituent quark masses through scalar fields ($\sigma$, $\zeta$, $\delta$) within an isospin asymmetric nuclear medium. 
\par Our findings have indicated that the effective quark masses of mesons decrease under the effect of the medium, suggesting partial restoration of chiral symmetry. Specifically, the effective mass of pseudoscalar $B$ mesons sharply decreases as baryonic density \(\rho_B\) increases. A similar trend is observed in vector $B$ mesons up to a critical baryonic density, after which the effective mass shows increasing behavior. This increase is likely due to the dominance of spin-dependent terms in the hyperfine interaction. 
The presence of isospin asymmetry results in the breaking of mass degeneracy between the ($B^{+}$, $B^{0}$) and ($B^{\ast+}$, $B^{\ast0}$) mesons, which induces contrasting behaviors in response to baryonic density. Furthermore, the increase in temperature of the medium results in a slower decrease in the effective masses of \(B\) mesons as a function of \(\rho_B\). 
We have also examined the weak decay constants of $B$ mesons in the nuclear medium by varying the \(\rho_B\), \(\eta\), and T. 
Both pseudoscalar and vector mesons experience a pronounced reduction in weak decay constants at smaller values of baryonic density. 
The influence of isospin asymmetry and temperature on the weak decay constants parallels that on the effective masses. The isospin-asymmetric nuclear medium is observed to impact the weak decay constants of vector $B$ mesons more moderately than those of pseudoscalar mesons, likely due to the modified invariant meson mass factor \(1/\textbf{\textit{M}}^{\ast}_{0}\). 
\par We have also calculated the DAs of both pseudoscalar ($B^{+}, B^{0}$) and vector ($B^{\ast+}, B^{\ast0}$) mesons in free space and within the isospin-asymmetric nuclear medium, for different values of \(\rho_B/\rho_0\),  \(\eta\), and \(T\). As \(\rho_B/\rho_0\) increases, the peaks of the DAs for the $B$ mesons shift to higher longitudinal momentum fractions (\(x\)). An increase in the isospin asymmetry of the medium leads to a splitting between the peaks of isospin doublets. 
Furthermore, an increase in the temperature of the medium results in a rise in the amplitude of the distribution functions for the $B$ mesons. In conclusion, our study has delivered a comprehensive evaluation of the characteristics of pseudoscalar and vector \(B\) mesons, offering vital insights into the behavior of $B$ mesons in a medium, which may have substantial consequences for the exploration of nuclear matter under extreme circumstances.

\section{Acknowledgement}
H.D. would like to thank  the Science and Engineering Research Board, Anusandhan-National Research Foundation, Government of India under the scheme SERB-POWER Fellowship (Ref No. SPF/2023/000116) for financial support.

\end{document}